\documentclass[journal]{IEEEtran}

\usepackage[
  colorlinks=true,
  linkcolor=blue,
  citecolor=blue,
  urlcolor=blue
]{hyperref}

\usepackage{pgfplots}
\usepackage[linesnumbered,ruled,lined]{algorithm2e}
\usetikzlibrary{shapes.multipart,intersections}
\usepackage{amsmath,amssymb,amsfonts,amsthm,steinmetz}
\usepackage{mathrsfs}  
\usepackage{textcomp}
\usepackage{acronym}
\usepackage{xcolor}
\usepackage{upgreek,xspace}
\usepackage{array}
\usepackage{tikz}
\usetikzlibrary{calc}
\makeatletter
\newcommand{\gettikzxy}[3]{%
  \tikz@scan@one@point\pgfutil@firstofone#1\relax
  \edef#2{\the\pgf@x}%
  \edef#3{\the\pgf@y}%
}
\makeatother

\usepackage{comment}

\usepackage[draft]{todonotes}   
\usepackage[T1]{fontenc}

\usepackage{esvect}

\usepackage{times}
\usepackage{bm}
\usepackage{stmaryrd}
\usepackage{babel}
\usepackage{graphics, graphicx}
\usepackage{gensymb}
\usepackage{cite}
\usepackage{enumitem}
\usepackage{url}

\usepackage{mathtools}

\begin{document}

\title{Prototype-Aware Fundamental Electromagnetic Limits on Wavefront Synthesis with Programmable Metasurfaces}

\author{Philipp~del~Hougne,~\IEEEmembership{Member,~IEEE}
\thanks{
P.~del~Hougne is with the Department of Electronics and Nanoengineering, Aalto University, 00076 Espoo, Finland and with Univ Rennes, CNRS, IETR - UMR 6164, F-35000, Rennes, France. (e-mail: philipp.del-hougne@univ-rennes.fr)
}
\thanks{This work was supported in part by the Nokia Foundation (project 20260028), the ANR France 2030 program (project ANR-22-PEFT-0005), the ANR PRCI program (project ANR-22-CE93-0010), the Rennes M\'etropole AES program (project ``SRI''), the European Union's European Regional Development Fund, and the French region of Brittany and Rennes Métropole through the contrats de plan État-Région program (projects ``SOPHIE/STIC \& Ondes'' and ``CyMoCoD'').}
}

\maketitle

\begin{abstract}
Wavefront synthesis is a central objective in many applications of programmable metasurfaces (PMs), ranging from electromagnetic holography and computational imaging to massive backscatter communications. Yet, fundamental limits on the ability of a given real-world PM prototype to synthesize a desired output wavefront remain largely unknown. Here, we derive prototype-aware and electromagnetically consistent bounds on target-wavefront synthesis in reconfigurable MIMO wave systems whose programmability stems from tunable lumped elements. Our approach combines multiport network theory (MNT), experimentally estimated proxy MNT parameters, and semidefinite relaxation. We account for relevant practical aspects of typical real-world PMs, such as mutual coupling, binary programmability, and lossy tunable loads. We derive bounds on strength-agnostic wavefront-synthesis fidelity, shape-agnostic target-mode strength, and the strength--fidelity Pareto frontier using two complementary threshold sweeps. We evaluate these bounds for four experimental MIMO systems whose transfer functions are parametrized by a reconfigurable intelligent surface (RIS), involving up to 100 1-bit-programmable elements and radio environments ranging from rich scattering to free space. Our bounds yield practical insights such as the identification of unattainable performance regions and the close-to-optimality certification of certain optimization outcomes. Comparisons with feasible discrete-optimization benchmarks show that the bounds can often be closely approached in practice, indicating tightness. While demonstrated with a RIS prototype, our methodology applies broadly to lumped-element-reconfigurable wave systems, including dynamic metasurface antennas.
Altogether, this work contributes to the development of a prototype-aware electromagnetic information theory for reconfigurable wave systems.
\end{abstract}


\section{Introduction}
\label{sec_introduction}

Programmable metasurfaces (PMs) such as reconfigurable intelligent surfaces (RISs) and dynamic metasurface antennas (DMAs) are emerging as technological enablers of wave-domain flexibility in next-generation wireless systems~\cite{pWDCperspective}. 
Many envisioned applications ultimately leverage the PM to manipulate the impinging field such that a desired scattered field is created. Prominent examples include programmable electromagnetic holograms~\cite{li2017electromagnetic}, the synthesis of specific illuminations for scene-aware computational imaging~\cite{li2019machine,del2020learned}, and the implementation of advanced modulation schemes in massive backscatter communications~\cite{zhao2020metasurface}. Consequently, understanding fundamental limits on the ability to synthesize a desired target scattered field with a given experimental PM prototype would be valuable. Yet, to the best of our knowledge, no prior work has tackled this problem.

More generally, despite a plethora of PM prototypes and demonstrations of PM-based functionalities, few works have explored the fundamental electromagnetic limits of PM-based systems. Importantly, since reconfigurability in the microwave regime almost always originates from tunable lumped elements (e.g., PIN diodes or varactors), most PM-based systems are amenable to an electromagnetically consistent description in terms of multiport network theory (MNT)~\cite{pWDCperspective}. Therein, each tunable lumped element is modeled as a ``virtual'' port terminated by a tunable load. 
The universality of this MNT-based system model implies that MNT-based techniques for evaluating fundamental limits of PM-based systems can be broadly applied to different PM embodiments such as RISs, DMAs, etc. Moreover, thanks to recent progress in \textit{experimentally} estimating very accurate proxy MNT parameters for PM-based system prototypes~\cite{sol2024experimentally,ContRIS_LWC,largeRIS_TCOM,del2025experimentalreducedrank,tapie2025experimental,tapie2026channel,del2026cross}, these fundamental limits can be evaluated for concrete experimental prototypes of PM-based systems. The accessible states of the tunable lumped elements in real-world PM prototypes are discrete (usually even binary) and typically \textit{not} lossless. 
Fundamental limits evaluated based on an experimentally calibrated MNT model simultaneously comply by construction with the laws of electromagnetism and a real-world feasibility set, which are jointly captured by the MNT model's mathematical structure and the experimentally estimated proxy MNT parameters. 

Recently, three types of prototype-aware fundamental electromagnetic limits of PM-based systems were derived and evaluated, namely for:
\begin{enumerate}[label=(\roman*)]
    \item the achievable multiplexing gain in PM-based backscatter multiple-input multiple-output (MIMO) systems~\cite{del2025effective},
    \item  the achievable channel gain and capacity in PM-assisted single-input single-output (SISO) systems~\cite{salmi2026electromagnetically},
    \item the achievable aggregate channel gain and operator-synthesis fidelity in PM-assisted MIMO systems~\cite{del2026electromagnetic}.
\end{enumerate}
Moreover, several related theoretical works obtained results under idealized feasibility sets, without prototype-aware grounding. Both (ii) and (iii) are based on a semidefinite relaxation (SDR) of a quadratically-constrained quadratic problem (QCQP) formulation of the objective.
Related to (ii), reactively loaded array pattern synthesis was formulated as a QCQP in~\cite{corcoles2015reactively}, and SDR-based fundamental bounds for pattern synthesis with reactively loaded antenna arrays were recently derived in~\cite{salmi2025optimization}, assuming continuously tunable, lossless lumped elements. Also related to (ii), a globally optimal solution for RIS-aided SISO channel gain maximization was derived in~\cite{wu2025beyond}, assuming that the ensemble of ``virtual'' ports associated with the RIS elements can be terminated by any lossless, reciprocal load network.

The notion of \textit{operator-synthesis fidelity} in (iii) refers to the ability to tune a reconfigurable MIMO system toward a desired linear input-output map (up to a complex-valued global prefactor). This \textit{operator-synthesis fidelity} generally differs from the \textit{wavefront-synthesis fidelity} discussed in the present paper; \textit{wavefront-synthesis fidelity} refers to the ability to jointly tune a reconfigurable MIMO system and its input wavefront such that the output wavefront approximates a desired one (up to a complex-valued global prefactor). Only in the special case of a single-input multiple-output (SIMO) system do \textit{operator-synthesis fidelity} and \textit{wavefront-synthesis fidelity} reduce to the same concept. Indeed, in the SIMO case, the input wavefront cannot be optimized and the output wavefront is proportional to the system's linear input-output map. Incidentally, for this special SIMO case, a so-called ``dense-controllability theorem'' was recently put forth in~\cite{dardari2026fundamental}. A key takeaway of this theorem is that if the Jacobian of the configuration-to-output-wavefront map has full rank at one configuration, then almost any normalized output wavefront can be approximated arbitrarily well by tuning the device configuration, assuming continuously tunable elements~\cite{dardari2026fundamental}.
For the general case of a reconfigurable MIMO system, no prototype-aware bounds on the \textit{wavefront-synthesis fidelity} have been derived or evaluated, to the best of our knowledge. 

While both~\cite{del2026electromagnetic} and~\cite{dardari2026fundamental} limit their analysis to the strength-agnostic fidelity (of the input-output map and of the output wavefront, respectively), in practical applications, not only the shape but also the strength (of the input-output map or output wavefront) matters. For instance, a closely approximated but very weak output wavefront may be useless or less useful than a random but strong output wavefront. There is typically a trade-off between fidelity and strength, because the available control in the reconfigurable system is limited. To capture this trade-off, prototype-aware fundamental limits on the strength--fidelity Pareto frontier would be very valuable. To the best of our knowledge, such limits on the Pareto frontier have been derived neither for operator synthesis nor for wavefront synthesis.

In this paper, we derive prototype-aware bounds on the strength--fidelity Pareto frontier for PM-based wavefront synthesis. Our contributions are summarized as follows.
\textit{First}, we derive fundamental prototype-aware and physics-consistent limits on wavefront-synthesis fidelity. The two key differences compared to~\cite{salmi2026electromagnetically,del2026electromagnetic} are that (i) the optimization objective concerns the output wavefront rather than the input-to-output map, and (ii) the optimization is over the system configuration and the input wavefront rather than only over the system configuration.
\textit{Second}, we derive fundamental prototype-aware and physics-consistent limits on the strength of the output wavefront. Again, the underlying optimization is over the system configuration and the input wavefront.
\textit{Third}, we derive fundamental prototype-aware bounds on the strength--fidelity Pareto frontier. This frontier quantifies the fundamental trade-off between accurately matching the desired wavefront shape and delivering substantial power into that wavefront. We compute outer bounds on this frontier using two complementary parameterizations: a fidelity-threshold sweep that maximizes target-mode strength subject to a minimum fidelity, and a strength-threshold sweep that maximizes fidelity subject to a minimum target-mode strength. \textit{Fourth}, we evaluate our bounds for four different experimental MIMO system prototypes involving an RIS with 100 1-bit-programmable elements. The 1-bit-programmability is typical of practical RIS prototypes and also found in the experiments in~\cite{li2017electromagnetic,li2019machine,zhao2020metasurface}. The four considered systems differ regarding the level of environmental scattering, ranging from rich scattering to free space.
Altogether, within the broader context of electromagnetic information theory (EIT)~\cite{di2024electromagnetic}, this paper contributes to the development of \textit{prototype-aware} EIT for \textit{reconfigurable} wave systems.

Our paper is organized as follows. In Sec.~\ref{sec_SystemModel}, we introduce the multiport-network system model. In Sec.~\ref{sec_Theory}, we derive SDR-based bounds on wavefront-synthesis fidelity, target-mode strength, and the strength--fidelity Pareto frontier. In Sec.~\ref{sec_discrete_optimizations}, we describe the feasible discrete-optimization benchmarks used to assess bound tightness. In Sec.~\ref{sec_ExpResults}, we evaluate the proposed bounds on four experimental RIS-parametrized MIMO systems. We close with a brief discussion in Sec.~\ref{sec_Discussion} and conclusion in Sec.~\ref{sec_Conclusion}.

\textit{Notation:}
$\mathbb{R}$, $\mathbb{C}$, and $\mathbb{B}\triangleq\{0,1\}$ denote the sets of real, complex, and binary numbers, respectively.
$\jmath\triangleq\sqrt{-1}$ denotes the imaginary unit.
$(\cdot)^*$ denotes elementwise complex conjugation.
$(\cdot)^\top$ and $(\cdot)^\dagger$ denote transpose and conjugate transpose, respectively.
$|\cdot|$ denotes absolute value.
$\|\cdot\|_2$ denotes the Euclidean norm of a vector.
$\mathbf{I}_a$ denotes the $a\times a$ identity matrix.
$\mathbf 0$ and $\mathbf 1$ denote the all-zeros and all-ones vectors/matrices of appropriate sizes.
$\mathrm{diag}(\mathbf{a})$ denotes the diagonal matrix whose diagonal entries are given by the vector $\mathbf{a}$.
$\mathrm{tr}(\cdot)$ denotes the trace operator. 
$\mathbf{A}\succeq\mathbf 0$ denotes that $\mathbf{A}$ is a positive semidefinite matrix.
$\mathbf{A}_{\mathcal{B}\mathcal{C}}$ denotes the block of $\mathbf{A}$ selected by row indices $\mathcal{B}$ and column indices $\mathcal{C}$.

\section{System Model}
\label{sec_SystemModel}

We consider, at a single target operating frequency, a linear wave system with $N_\mathrm{T}$ input ports and $N_\mathrm{R}$ output ports. The system's reconfigurability stems from $N_\mathrm{S}$ tunable lumped elements. We model each tunable lumped element as a ``virtual'' port terminated by a tunable load. We partition the reconfigurable system into its static and tunable components. The static subsystem has $N=N_\mathrm{T}+N_\mathrm{R}+N_\mathrm{S}$ ports and is characterized by its scattering matrix $\mathbf{S}\in\mathbb{C}^{N\times N}$. The tunable subsystem is the ensemble of the $N_\mathrm{S}$ tunable loads; it has $N_\mathrm{S}$ ports and is characterized by its scattering matrix $\mathbf{\Phi}\in\mathbb{C}^{N_\mathrm{S}\times N_\mathrm{S}}$, which is by construction diagonal and given by $\mathbf{\Phi}(\mathbf{r})=\mathrm{diag}(\mathbf{r})$, where $\mathbf{r}=[r_1,\ldots,r_{N_\mathrm S}]^\top\in\mathbb{C}^{N_\mathrm S}$ is the load vector and $r_i$ denotes the reflection coefficient of the $i$th load. In line with our PM prototype based on PIN diodes, we assume that $r_i\in\{\alpha,\beta\}$, where $\alpha\in\mathbb{C}$ and $\beta\in\mathbb{C}$ are the two possible reflection coefficients of a tunable load associated with a PIN diode.

The system's end-to-end MIMO transfer function $\mathbf{H}\in\mathbb{C}^{N_\mathrm{R}\times N_\mathrm{T}}$ (also referred to as the end-to-end MIMO channel matrix) can be evaluated based on the connection of the static $N$-port subsystem and the tunable $N_\mathrm{S}$-port subsystem via the $N_\mathrm{S}$ ``virtual'' ports. Standard MNT\footnote{A detailed derivation can be found, for instance, in Appendix~B.3 of~\cite{prod2024updatable}.} yields
\begin{equation}
\mathbf{H}(\mathbf{r}) = \mathbf{H}_0 + \mathbf{A}\,\bigl(\mathbf{I}_{N_\mathrm{S}}\,-\,\mathbf{\Phi}(\mathbf{r})\,\mathbf{\Gamma}\bigr)^{-1}\,\mathbf{\Phi}(\mathbf{r})\,\mathbf{B},
\label{eq_MNT}
\end{equation}
where, for notational ease, $\mathbf{H}_0 \triangleq \mathbf{S}_\mathcal{RT}\in\mathbb{C}^{N_\mathrm{R}\times N_\mathrm{T}}$, $\mathbf{A} \triangleq \mathbf{S}_\mathcal{RS}\in\mathbb{C}^{N_\mathrm{R}\times N_\mathrm{S}}$, $\mathbf{\Gamma} \triangleq \mathbf{S}_\mathcal{SS}\in\mathbb{C}^{N_\mathrm{S}\times N_\mathrm{S}}$, and $\mathbf{B} \triangleq \mathbf{S}_\mathcal{ST}\in\mathbb{C}^{N_\mathrm{S}\times N_\mathrm{T}}$, and $\mathcal{T}$, $\mathcal{R}$, and $\mathcal{S}$ denote the sets containing the port indices associated with the transmitting ports, the receiving ports, and the ``virtual'' ports, respectively.  Throughout this paper, we assume that $\mathbf I_{N_\mathrm S}-\mathbf\Phi(\mathbf r)\,\mathbf\Gamma$ is nonsingular for every admissible binary load vector $\mathbf r$. 

The system's output wavefront $\mathbf{y}\in\mathbb{C}^{N_\mathrm{R}}$ depends linearly on the system's input wavefront $\mathbf{x}\in\mathbb{C}^{N_\mathrm{T}}$ and, as per \eqref{eq_MNT}, non-linearly on the system's load vector $\mathbf{r}$:
\begin{equation}
    \mathbf{y} = \mathbf{H}(\mathbf{r})\,\mathbf{x},
\label{eq_def_y}
\end{equation}
where we impose the input-power normalization $\|\mathbf x\|_2^2\le 1$. 
Since $\mathbf{x}$ and $\mathbf{y}$ are defined as power-wave amplitudes at physical ports and the overall system is passive, the total outgoing power over all physical ports cannot exceed the incident power, implying $\|\mathbf y\|_2^2\le \|\mathbf x\|_2^2$.

\section{Bounds on Fidelity and Strength in\\ PM-Based Target-Wavefront Synthesis}
\label{sec_Theory}

In this section, we derive SDR-based bounds on the wavefront-synthesis performance of PM-based systems. We first introduce fidelity and target-mode strength as the two performance metrics of interest in Sec.~\ref{subsec_Metrics}. We then reformulate the MNT input-output relation in Sec.~\ref{subsec_AuxVariable} by introducing an auxiliary internal-state variable, which makes the output wavefront linear in an augmented variable collecting the internal state and the input wavefront. The binary programmability constraints are subsequently expressed in Sec.~\ref{subsec_binary_constraints} as homogeneous quadratic equalities in this augmented variable. This reformulation enables us to cast the relevant performance-limit problems as QCQPs or fractional QCQPs. We then lift the augmented variable to a positive semidefinite matrix and drop the resulting rank-one constraint, yielding convex semidefinite programs (SDPs) whose optimal values upper-bound the corresponding achievable performance~\cite{boyd2004convex,luo2010semidefinite}. Specifically, we derive SDR-based bounds on strength-agnostic fidelity in Sec.~\ref{subsec_shape_only}, shape-agnostic target-mode strength in Sec.~\ref{subsec_strength_only}, and the strength--fidelity Pareto frontier using two complementary threshold sweeps in Secs.~\ref{subsec_smin_sweep} and~\ref{subsec_fmin_sweep}.

\subsection{Fidelity and Strength Metrics}
\label{subsec_Metrics}

We denote by $\mathbf y_\star\in\mathbb{C}^{N_\mathrm R}$ the desired nonzero output wavefront, and moreover we define $\hat{\mathbf y}_\star=\mathbf y_\star / \|\mathbf y_\star\|_2$. 
Then, we further define the target-mode projector $\mathbf P_\star=\hat{\mathbf y}_\star\hat{\mathbf y}_\star^\dagger$ and its orthogonal complement $\mathbf P_\perp = \mathbf I_{N_\mathrm R}-\mathbf P_\star$. For a given output wavefront $\mathbf{y}$, we can now evaluate its target-mode strength $S$ and leakage power $L$:
\begin{align}
    S(\mathbf y,\mathbf y_\star) &=\|\mathbf P_\star\mathbf y\|_2^2
    =|\hat{\mathbf y}_\star^\dagger\mathbf y|^2,
    \label{eq_strength}\\
    L(\mathbf y,\mathbf y_\star) &=\|\mathbf P_\perp\mathbf y\|_2^2.
    \label{eq_leakage}
\end{align}
Moreover, we can express the fidelity $F$ of $\mathbf{y}$ with respect to $\mathbf y_\star $ in terms of $S$ and $L$:
\begin{equation}
    F(\mathbf y,\mathbf y_\star)=\frac{|\mathbf y_\star^\dagger \mathbf y|^2}{\|\mathbf y_\star\|_2^2\|\mathbf y\|_2^2}
     =\frac{S(\mathbf y,\mathbf y_\star)}{S(\mathbf y,\mathbf y_\star)+L(\mathbf y,\mathbf y_\star)},
    \label{eq_F_SL}
\end{equation}
provided $S+L>0$. $F$ is by construction invariant to a global complex scalar multiplying $\mathbf y$ but remains sensitive to relative amplitudes and relative phases across the entries of $\mathbf{y}$ and $\mathbf y_\star $. Moreover, \(F\leq1\) by definition, and fidelity is evaluated only for nonzero output wavefronts.

For a fixed admissible load vector \(\mathbf r\), unit fidelity is achievable for any desired output wavefront if \(\operatorname{rank}(\mathbf H(\mathbf r))=N_\mathrm R\). This condition requires \(N_\mathrm T\ge N_\mathrm R\) and means that the input wavefront has enough degrees of freedom to synthesize an arbitrary output wavefront up to a nonzero complex scalar. Hence, in such cases, the strength-agnostic fidelity problem becomes trivial: one can always choose an input wavefront \(\mathbf x\) such that \(\mathbf H(\mathbf r)\mathbf x=\kappa\,\mathbf y_\star\) for some \(\kappa\in\mathbb C\setminus\{0\}\).

\subsection{Auxiliary Variable}
\label{subsec_AuxVariable}

We introduce the auxiliary variable
\begin{equation}
\mathbf v \triangleq (\mathbf I_{N_\mathrm S}-\mathbf\Phi(\mathbf r)\mathbf\Gamma)^{-1}
\mathbf\Phi(\mathbf r)\,\mathbf B\,\mathbf x \in\mathbb{C}^{N_\mathrm S}
\label{eq_internal_state}
\end{equation}
that allows us to combine \eqref{eq_MNT} and \eqref{eq_def_y} as follows: 
\begin{align}
\mathbf y &=\mathbf H_0\,\mathbf x+\mathbf A\,\mathbf v,\\
\mathbf y &=\mathbf C\,\bm\xi,
\label{eq_y_linear_xv}
\end{align}
where we introduce $\mathbf C \triangleq \left[ \mathbf{A} \ \mathbf H_0 \right] \in\mathbb{C}^{N_\mathrm R\times (N_\mathrm S+N_\mathrm T)}$ and $\bm\xi \triangleq  \left[\mathbf v^\top \ \mathbf x^\top\right]^\top \in\mathbb{C}^{N_\mathrm S+N_\mathrm T}$.

\subsection{Binary Programmability as Quadratic Constraints}
\label{subsec_binary_constraints}

We can rewrite  \eqref{eq_internal_state} as $\left(\mathbf I_{N_\mathrm S}-\mathbf\Phi(\mathbf r)\,\mathbf\Gamma\right)\,\mathbf v
=
\mathbf\Phi(\mathbf r)\,\mathbf B\,\mathbf x$ and further as
\begin{align}
\mathbf v &=\mathbf\Phi(\mathbf r)\left(\mathbf\Gamma\mathbf v+\mathbf B\mathbf x\right),\\
\mathbf v &=\mathbf\Phi(\mathbf r) \, \mathbf z,
\label{eq_v_phi_z}
\end{align}
where we introduce $\mathbf z \triangleq \mathbf\Gamma\,\mathbf v+\mathbf B\,\mathbf x$.
The $i$th tunable load imposes
\begin{equation}
v_i = r_i \, z_i,
\qquad r_i\in\{\alpha,\beta\}.
\label{eq_vm_rm_zm}
\end{equation}
Analogous to~\cite{shim2024fundamental,gertler2025many,salmi2026electromagnetically,del2026electromagnetic}, we note that the logical binary condition in \eqref{eq_vm_rm_zm} is equivalent to
\begin{equation}
(v_i-\alpha \, z_i)^*\, (v_i-\beta\,  z_i)=0.
\label{eq_binary_or}
\end{equation}

We now express \eqref{eq_binary_or} as a quadratic equality in \(\bm\xi\). Our starting point is 
\begin{equation}
    z_i=\bm{\gamma}_i^\top\mathbf v+\mathbf b_i^\top\mathbf x,
\end{equation}
where \(\bm{\gamma}_i^\top\) denotes the \(i\)th row of \(\mathbf{\Gamma}\) and \(\mathbf{b}_i^\top\) denotes the \(i\)th row of \(\mathbf{B}\).
Next, we express the factor \(v_i-\rho \,z_i\), where \(\rho\in\{\alpha,\beta\}\), as a linear function of \(\bm\xi\):
\begin{equation}
    v_i-\rho\, z_i
    =
    \mathbf g_{i,\rho}^\top\bm\xi,
    \label{eq_linear_factor}
\end{equation}
where  
\begin{equation}
    \mathbf g_{i,\rho}
    \triangleq
    \begin{bmatrix}
        \mathbf e_i-\rho\,\bm{\gamma}_i\\
        -\rho\,\mathbf b_i
    \end{bmatrix}
    \in\mathbb{C}^{N_\mathrm S+N_\mathrm T} 
    \label{eq_g_i_rho}
\end{equation}
and \(\mathbf{e}_i\) denotes the \(i\)th canonical basis vector in \(\mathbb{C}^{N_\mathrm S}\).

Substituting \eqref{eq_linear_factor} into \eqref{eq_binary_or} yields
\begin{equation}
    \left(\mathbf g_{i,\alpha}^\top\,\bm\xi\right)^*
    \left(\mathbf g_{i,\beta}^\top\,\bm\xi\right)=0.
    \label{eq_binary_g}
\end{equation}
Equivalently,
\begin{equation}
    \bm\xi^\dagger\,\mathbf R_i\,\bm\xi=0,
    \label{eq_Ri_def}
\end{equation}
where $\mathbf R_i\triangleq \mathbf g_{i,\alpha}^*\mathbf g_{i,\beta}^\top$.

The binary programmability constraints of the $N_\mathrm{S}$ tunable lumped elements can thus be expressed as $N_\mathrm{S}$ complex-valued quadratic constraints of the form \eqref{eq_Ri_def} with \(i=1,\ldots,N_\mathrm S\).

\subsection{Strength-Agnostic Fidelity Bound}
\label{subsec_shape_only}

In this subsection, we derive a bound on the largest achievable fidelity, irrespective of the delivered target-mode strength. The underlying optimization problem is
\begin{equation}
\begin{aligned}
\max_{\mathbf r,\mathbf{x}}\quad 
& F \left(\mathbf y(\mathbf{r},\mathbf{x}),\mathbf y_\star\right)\\
\text{s.t.}\quad 
& r_i\in\{\alpha,\beta\},\quad i=1,\dots,N_\mathrm S.
\end{aligned}
\label{eq:fidelity_opt}
\end{equation}
We do not include the power constraint on $\mathbf{x}$ in \eqref{eq:fidelity_opt} because the objective in \eqref{eq:fidelity_opt} is scale invariant. 
Defining 
\begin{subequations}
\begin{align}
    \mathbf Q_\mathrm{S} &\triangleq \mathbf C^\dagger\,\mathbf P_\star\,\mathbf C,\\
    \mathbf Q_\mathrm{L} &\triangleq \mathbf C^\dagger\,\mathbf P_\perp\,\mathbf C,\\
    \mathbf Q_\mathrm{D} &\triangleq \mathbf Q_\mathrm{S}+\mathbf Q_\mathrm{L}=\mathbf C^\dagger\,\mathbf C,
\end{align}
\label{eq_QS_QL}
\end{subequations}
we can rewrite \eqref{eq:fidelity_opt} as  
\begin{equation}
\begin{aligned}
\max_{\bm\xi}\quad
&\frac{\bm\xi^\dagger\, \mathbf Q_\mathrm{S}\, \bm\xi}{\bm\xi^\dagger\, \mathbf Q_\mathrm{D}\, \bm\xi}\\
\mathrm{s.t.}\quad
&\bm\xi^\dagger\, \mathbf R_i\, \bm\xi=0,
\quad i=1,\ldots,N_\mathrm S,
\end{aligned}
\label{eq_shape_qcqp_latest}
\end{equation}
where we use $S= \bm\xi^\dagger\,\mathbf Q_\mathrm{S}\,\bm\xi$ and $S+L=\bm\xi^\dagger\,\mathbf Q_\mathrm{D}\,\bm\xi$.

The objective in \eqref{eq_shape_qcqp_latest} is the ratio of two functions that are each quadratic in $\bm\xi$, implying that \eqref{eq_shape_qcqp_latest} is not a standard QCQP.
To obtain a problem formulation that is suitable for SDR, we use a Charnes--Cooper normalization~\cite{charnes1962programming,boyd2004convex}. The same idea was used in~\cite{del2026electromagnetic} to handle the fractional-quadratic operator-synthesis fidelity objective. In the present shape-only wavefront-synthesis problem, the transformation is simpler because both numerator and denominator of the wavefront-synthesis fidelity objective are homogeneous quadratic forms in \(\bm\xi\), and all binary-programmability constraints in \eqref{eq_Ri_def} are homogeneous as well.

For any feasible nonzero \(\bm\xi\) with $d(\bm\xi)\triangleq \bm\xi^\dagger\,\mathbf Q_\mathrm{D}\,\bm\xi>0$, we can rescale \(\bm\xi\) without changing either the fidelity objective or the binary constraints. We thus define the normalized variable
\begin{equation}
    \tilde{\bm\xi}\triangleq \frac{\bm\xi}{\sqrt{d(\bm\xi)}},
    \label{eq_xi_tilde_def}
\end{equation}
for which $\tilde{\bm\xi}^{\dagger}\,\mathbf Q_\mathrm{D}\,\tilde{\bm\xi}=1$ by construction. Now, we can rewrite \eqref{eq_shape_qcqp_latest} as 
\begin{equation}
\begin{aligned}
\max_{ \tilde{\bm\xi}}\quad
& \tilde{\bm\xi}^\dagger\,\mathbf Q_\mathrm{S}\, \tilde{\bm\xi}\\
\mathrm{s.t.}\quad
& \tilde{\bm\xi}^\dagger\,\mathbf Q_\mathrm{D}\, \tilde{\bm\xi}=1,\\
& \tilde{\bm\xi}^\dagger\,\mathbf R_i\, \tilde{\bm\xi}=0,
\quad i=1,\ldots,N_\mathrm S.
\end{aligned}
\label{eq_shape_qcqp_normalized}
\end{equation}

We now lift the problem in \eqref{eq_shape_qcqp_normalized} by defining
\begin{equation}
    \tilde{\mathbf \Xi} \triangleq \tilde{\bm\xi}\,\tilde{\bm\xi}^\dagger
    \in \mathbb C^{(N_\mathrm S+N_\mathrm T)\times(N_\mathrm S+N_\mathrm T)}.
    \label{eq_Xi_lift}
\end{equation}
Then, \(\tilde{\mathbf \Xi}\succeq\mathbf 0\) and \(\mathrm{rank}(\tilde{\mathbf \Xi})=1\), and each quadratic form can be written as $\tilde{\bm\xi}^\dagger\,\mathbf P\,\tilde{\bm\xi} = \mathrm{tr}(\mathbf P\,\tilde{\mathbf \Xi})$ with $\mathbf P\in\{\mathbf Q_\mathrm{S},\mathbf Q_\mathrm{D},\mathbf R_1,\ldots,\mathbf R_{N_\mathrm S}\}$. 
The exact lifted formulation of  \eqref{eq_shape_qcqp_normalized} is therefore
\begin{equation}
\begin{aligned}
\max_{\tilde{\mathbf \Xi}}\quad
&\mathrm{tr}(\mathbf Q_\mathrm{S}\,\tilde{\mathbf \Xi})\\
\mathrm{s.t.}\quad
&\mathrm{tr}(\mathbf Q_\mathrm{D}\,\tilde{\mathbf \Xi})=1,\\
&\mathrm{tr}(\mathbf R_i\,\tilde{\mathbf \Xi})=0,
\quad i=1,\ldots,N_\mathrm S,\\
&\tilde{\mathbf \Xi}\succeq\mathbf0,\\
&\mathrm{rank}(\tilde{\mathbf \Xi})=1.
\end{aligned}
\label{eq_shape_rank_lifted}
\end{equation}
The only nonconvex constraint in \eqref{eq_shape_rank_lifted} is the rank-one constraint.
The constraints involving $\mathbf R_i$ are generally complex-valued linear equalities and are understood throughout this paper as imposing equality of both their real and imaginary parts.
Dropping the rank-one constraint in \eqref{eq_shape_rank_lifted} yields the SDR
\begin{equation}
\begin{aligned}
\max_{\tilde{\mathbf \Xi}}\quad
&\mathrm{tr}(\mathbf Q_\mathrm{S}\,\tilde{\mathbf \Xi})\\
\mathrm{s.t.}\quad
&\mathrm{tr}(\mathbf Q_\mathrm{D}\,\tilde{\mathbf \Xi})=1,\\
&\mathrm{tr}(\mathbf R_i\,\tilde{\mathbf \Xi})=0,
\quad i=1,\ldots,N_\mathrm S,\\
&\tilde{\mathbf \Xi}\succeq\mathbf0.
\end{aligned}
\label{eq_shape_sdr}
\end{equation}
The problem in \eqref{eq_shape_sdr} is a convex SDP and can be solved with standard convex optimization solvers. Because every feasible point of the exact rank-constrained formulation in \eqref{eq_shape_rank_lifted} is feasible for the relaxed SDP in \eqref{eq_shape_sdr}, the optimal value of \eqref{eq_shape_sdr} is an upper bound on the largest achievable shape-only wavefront-synthesis fidelity. As demonstrated in Appendix~\ref{app_ambiguity_all}, this bound is agnostic to inevitable ambiguities in experimentally estimated proxy MNT parameters.

\subsection{Shape-Agnostic Strength Bound}
\label{subsec_strength_only}

In this subsection, we derive a bound on the largest achievable target-mode strength, irrespective of the corresponding shape. The underlying optimization problem is
\begin{equation}
\begin{aligned}
\max_{\mathbf r,\mathbf{x}}\quad 
& S \left(\mathbf y(\mathbf{r},\mathbf{x}),\mathbf y_\star\right)\\
\text{s.t.}\quad 
& \|\mathbf x\|_2^2\le 1,\\
& r_i\in\{\alpha,\beta\},\quad i=1,\dots,N_\mathrm S.
\end{aligned}
\label{eq:strength_opt}
\end{equation}

With $\mathbf E \triangleq \begin{bmatrix}
    \mathbf0 & \mathbf0\\
    \mathbf0 & \mathbf I_{N_\mathrm T}
    \end{bmatrix}$, 
we can write the power constraint on $\mathbf{x}$ as $\bm\xi^\dagger\,\mathbf E\,\bm\xi\le 1$ and recast \eqref{eq:strength_opt} as the following QCQP:
\begin{equation}
\begin{aligned}
\max_{\bm\xi}\quad
&\bm\xi^\dagger\,\mathbf Q_\mathrm{S}\,\bm\xi\\
\mathrm{s.t.}\quad
&\bm\xi^\dagger\,\mathbf E\,\bm\xi\le1,\\
&\bm\xi^\dagger\,\mathbf R_i\,\bm\xi=0,
\quad i=1,\ldots,N_\mathrm S.
\end{aligned}
\label{eq_strength_only_qcqp}
\end{equation}
Lifting \eqref{eq_strength_only_qcqp} with
\begin{equation}
    \mathbf\Xi \triangleq \bm\xi\,\bm\xi^\dagger
    \in \mathbb C^{(N_\mathrm S+N_\mathrm T)\times(N_\mathrm S+N_\mathrm T)}
    \label{eq_Xi_lift_strength}
\end{equation}
gives the exact rank-constrained formulation
\begin{equation}
\begin{aligned}
\max_{\mathbf\Xi}\quad
&\mathrm{tr}(\mathbf Q_\mathrm{S}\,\mathbf\Xi)\\
\mathrm{s.t.}\quad
&\mathrm{tr}(\mathbf E\,\mathbf\Xi)\le1,\\
&\mathrm{tr}(\mathbf R_i\,\mathbf\Xi)=0,
\quad i=1,\ldots,N_\mathrm S,\\
&\mathbf\Xi\succeq\mathbf0,\\
&\mathrm{rank}(\mathbf\Xi)=1.
\end{aligned}
\label{eq_strength_only_rank_lifted}
\end{equation}
As noted above, the constraints involving $\mathbf R_i$ are understood as complex linear equalities. The only nonconvex constraint in \eqref{eq_strength_only_rank_lifted} is the rank-one constraint. Dropping it yields the SDR
\begin{equation}
\begin{aligned}
\max_{\mathbf\Xi}\quad
&\mathrm{tr}(\mathbf Q_\mathrm{S}\,\mathbf\Xi)\\
\mathrm{s.t.}\quad
&\mathrm{tr}(\mathbf E\,\mathbf\Xi)\le1,\\
&\mathrm{tr}(\mathbf R_i\,\mathbf\Xi)=0,
\quad i=1,\ldots,N_\mathrm S,\\
&\mathbf\Xi\succeq\mathbf0.
\end{aligned}
\label{eq_strength_only_sdr}
\end{equation}
The problem in \eqref{eq_strength_only_sdr} is a convex SDP and can be solved with standard convex optimization solvers. Because every feasible point of the exact rank-constrained formulation in \eqref{eq_strength_only_rank_lifted} is feasible for the relaxed SDP in \eqref{eq_strength_only_sdr}, the optimal value of \eqref{eq_strength_only_sdr} is an upper bound on the largest achievable shape-agnostic target-mode strength. As demonstrated in Appendix~\ref{app_ambiguity_all}, this bound is agnostic to inevitable ambiguities in experimentally estimated proxy MNT parameters.

\subsection{Strength--Fidelity Pareto Frontier Bound from Strength-Threshold Sweep}
\label{subsec_smin_sweep}

To determine a bound on the strength--fidelity Pareto frontier, we consider the following fidelity-maximization problem subject to a minimum-strength constraint:
\begin{equation}
\begin{aligned}
\max_{\bm\xi}\quad
&\frac{\bm\xi^\dagger\,\mathbf Q_\mathrm{S}\,\bm\xi}{\bm\xi^\dagger\,\mathbf Q_\mathrm{D}\,\bm\xi}\\
\mathrm{s.t.}\quad
&\bm\xi^\dagger\,\mathbf Q_\mathrm{S}\,\bm\xi\ge S_\mathrm{min},\\
&\bm\xi^\dagger\,\mathbf E\,\bm\xi\le1,\\
&\bm\xi^\dagger\,\mathbf R_i\,\bm\xi=0,
\quad i=1,\ldots,N_\mathrm S.
\end{aligned}
\label{eq_Smin_qcqp}
\end{equation}
The only differences between \eqref{eq_Smin_qcqp} and \eqref{eq_shape_qcqp_latest} are the minimum-strength constraint and the power constraint (that is necessary for a meaningful minimum-strength constraint) in the former. 
Similar to \eqref{eq_shape_qcqp_latest}, the objective in \eqref{eq_Smin_qcqp} is a fraction of two functions that are each quadratic in $\bm\xi$. Unlike \eqref{eq_shape_qcqp_latest}, \eqref{eq_Smin_qcqp} is not scale-invariant and thus nonhomogeneous. Indeed, the minimum-strength constraint and the input-power constraint set an absolute scale for \(\bm\xi\).
To obtain a problem formulation suitable for SDR, we use a Charnes--Cooper transformation with an explicit scaling variable~\cite{charnes1962programming,boyd2004convex}.

For any feasible nonzero \(\bm\xi\) with \(\bm\xi^\dagger\,\mathbf Q_\mathrm{D}\,\bm\xi>0\), we define the scaled variable
\begin{equation}
    \breve{\bm\xi}\triangleq \sqrt{t}\,\bm\xi,
    \label{eq_define_brevexi}
\end{equation}
where
\begin{equation}
    t\triangleq\frac{1}{\bm\xi^\dagger\,\mathbf Q_\mathrm{D}\,\bm\xi},
    \label{eq_define_t_for_brevexi}
\end{equation}
such that \(\breve{\bm\xi}^\dagger \,\mathbf Q_\mathrm{D}\,\breve{\bm\xi}=1\). 
Now, we can equivalently write \eqref{eq_Smin_qcqp} in the scaled variables as
\begin{equation}
\begin{aligned}
\max_{\breve{\bm\xi},t}\quad
&\breve{\bm\xi}^\dagger\,\mathbf Q_\mathrm{S}\,\breve{\bm\xi}\\
\mathrm{s.t.}\quad
&\breve{\bm\xi}^\dagger\,\mathbf Q_\mathrm{D}\,\breve{\bm\xi}=1,\\
&\breve{\bm\xi}^\dagger\,\mathbf Q_\mathrm{S}\,\breve{\bm\xi}\ge S_\mathrm{min}t,\\
&\breve{\bm\xi}^\dagger\,\mathbf E\,\breve{\bm\xi}\le t,\\
&\breve{\bm\xi}^\dagger\,\mathbf R_i\,\breve{\bm\xi}=0,
\quad i=1,\ldots,N_\mathrm S,\\
&t>0.
\end{aligned}
\label{eq_Smin_CC_vector}
\end{equation}

Next, we define
\begin{equation}
    \breve{\mathbf\Xi}\triangleq \breve{\bm\xi}\,\breve{\bm\xi}^\dagger
    =t\,\bm\xi\,\bm\xi^\dagger,
    \label{eq_CC_smin}
\end{equation}
which allows us to formulate the following exact lifted version of \eqref{eq_Smin_CC_vector}:
\begin{equation}
\begin{aligned}
\max_{\breve{\mathbf\Xi},t}\quad
&\mathrm{tr}(\mathbf Q_\mathrm{S}\,\breve{\mathbf\Xi})\\
\mathrm{s.t.}\quad
&\mathrm{tr}(\mathbf Q_\mathrm{D}\,\breve{\mathbf\Xi})=1,\\
&\mathrm{tr}(\mathbf Q_\mathrm{S}\,\breve{\mathbf\Xi})\ge S_\mathrm{min}t,\\
&\mathrm{tr}(\mathbf E\,\breve{\mathbf\Xi})\le t,\\
&\mathrm{tr}(\mathbf R_i\,\breve{\mathbf\Xi})=0,
\quad i=1,\ldots,N_\mathrm S,\\
&\breve{\mathbf\Xi}\succeq\mathbf0,\\
& t>0,\\
&\mathrm{rank}(\breve{\mathbf\Xi})=1.
\end{aligned}
\label{eq_Smin_rank}
\end{equation}
As noted above, the constraints involving $\mathbf R_i$ are understood as complex linear equalities. The only nonconvex constraint in \eqref{eq_Smin_rank} is the rank-one constraint. 
Dropping it and replacing \(t>0\) by its closed convex relaxation \(t\ge0\) yields
\begin{equation}
\begin{aligned}
\max_{\breve{\mathbf\Xi},t}\quad
&\mathrm{tr}(\mathbf Q_\mathrm{S}\,\breve{\mathbf\Xi})\\
\mathrm{s.t.}\quad
&\mathrm{tr}(\mathbf Q_\mathrm{D}\,\breve{\mathbf\Xi})=1,\\
&\mathrm{tr}(\mathbf Q_\mathrm{S}\,\breve{\mathbf\Xi})\ge S_\mathrm{min}t,\\
&\mathrm{tr}(\mathbf E\,\breve{\mathbf\Xi})\le t,\\
&\mathrm{tr}(\mathbf R_i\,\breve{\mathbf\Xi})=0,
\quad i=1,\ldots,N_\mathrm S,\\
&\breve{\mathbf\Xi}\succeq\mathbf0,\\
&t\ge0.
\end{aligned}
\label{eq_Smin_sdr}
\end{equation}
The problem in \eqref{eq_Smin_sdr} is a convex SDP and can be solved with standard convex optimization solvers. Because every feasible point of the exact rank-constrained formulation in \eqref{eq_Smin_rank} is feasible for the relaxed SDP in \eqref{eq_Smin_sdr}, the optimal value of \eqref{eq_Smin_sdr} is an upper bound on the largest achievable fidelity among output wavefronts whose target-mode strength is at least \(S_\mathrm{min}\). As demonstrated in Appendix~\ref{app_ambiguity_all}, this bound is agnostic to inevitable ambiguities in experimentally estimated proxy MNT parameters.

To trace out the bound on the strength--fidelity Pareto frontier, we sweep through a range of $S_\mathrm{min}$ values and solve the corresponding \eqref{eq_Smin_sdr} for each of them. For the special case of \(S_\mathrm{min}=0\), the minimum-strength constraint is effectively inactive. Because the fidelity objective and the binary constraints are scale-invariant, the input-power constraint does not change the optimal value. Thus, the underlying problem reduces to the strength-agnostic fidelity problem in \eqref{eq_shape_qcqp_latest}.

\subsection{Strength--Fidelity Pareto Frontier Bound from Fidelity-Threshold Sweep}
\label{subsec_fmin_sweep}

As an alternative way of determining a bound on the strength--fidelity Pareto frontier, we consider, for $F_\mathrm{min}\in[0,1]$, the following strength-maximization problem subject to a minimum-fidelity constraint:
\begin{equation}
\begin{aligned}
\max_{\bm\xi}\quad
&\bm\xi^\dagger\,\mathbf Q_\mathrm{S}\,\bm\xi\\
\mathrm{s.t.}\quad
&\frac{\bm\xi^\dagger\,\mathbf Q_\mathrm{S}\,\bm\xi}{\bm\xi^\dagger\,\mathbf Q_\mathrm{D}\,\bm\xi}
\ge F_\mathrm{min},\\
&\bm\xi^\dagger\,\mathbf E\,\bm\xi\le1,\\
&\bm\xi^\dagger\,\mathbf R_i\,\bm\xi=0,
\quad i=1,\ldots,N_\mathrm S.
\end{aligned}
\label{eq_Fmin_qcqp}
\end{equation}
The only difference between \eqref{eq_Fmin_qcqp} and \eqref{eq_strength_only_qcqp} is the minimum-fidelity constraint in the former. 
For $F_\mathrm{min}=0$, this constraint is effectively inactive and \eqref{eq_Fmin_qcqp} reduces to \eqref{eq_strength_only_qcqp}.
For $F_\mathrm{min}>0$, the minimum-fidelity constraint is equivalently
\begin{equation}
    (1-F_\mathrm{min})\,\bm\xi^\dagger\,\mathbf Q_\mathrm{S}\,\bm\xi
    -F_\mathrm{min}\,\bm\xi^\dagger\mathbf Q_\mathrm{L}\,\bm\xi\ge0.
    \label{eq_Fmin_linearized_qc}
\end{equation}

Lifting \eqref{eq_Fmin_qcqp} with
\begin{equation}
    \mathbf\Xi\triangleq\bm\xi\,\bm\xi^\dagger
    \in \mathbb C^{(N_\mathrm S+N_\mathrm T)\times(N_\mathrm S+N_\mathrm T)}
    \label{eq_Xi_lift_Fmin}
\end{equation}
allows us to formulate the following exact lifted version of \eqref{eq_Fmin_qcqp}:
\begin{equation}
\begin{aligned}
\max_{\mathbf\Xi}\quad
&\mathrm{tr}(\mathbf Q_\mathrm{S}\,\mathbf\Xi)\\
\mathrm{s.t.}\quad
&(1-F_\mathrm{min})\,\mathrm{tr}(\mathbf Q_\mathrm{S}\,\mathbf\Xi)
-F_\mathrm{min}\,\mathrm{tr}(\mathbf Q_\mathrm{L}\,\mathbf\Xi)\ge0,\\
&\mathrm{tr}(\mathbf E\,\mathbf\Xi)\le1,\\
&\mathrm{tr}(\mathbf R_i\,\mathbf\Xi)=0,
\quad i=1,\ldots,N_\mathrm S,\\
&\mathbf\Xi\succeq\mathbf0,\\
&\mathrm{rank}(\mathbf\Xi)=1.
\end{aligned}
\label{eq_Fmin_rank}
\end{equation}
As noted above, the constraints involving $\mathbf R_i$ are understood as complex linear equalities. The only nonconvex constraint in \eqref{eq_Fmin_rank} is the rank-one constraint. Dropping it yields the SDR
\begin{equation}
\begin{aligned}
\max_{\bm\Xi}\quad
&\mathrm{tr}(\mathbf Q_\mathrm{S}\,\mathbf \Xi)\\
\mathrm{s.t.}\quad
&(1-F_\mathrm{min})\,\mathrm{tr}(\mathbf Q_\mathrm{S}\,\mathbf\Xi)
-F_\mathrm{min}\,\mathrm{tr}(\mathbf Q_\mathrm{L}\,\mathbf \Xi)\ge0,\\
&\mathrm{tr}(\mathbf E\,\mathbf \Xi)\le1,\\
&\mathrm{tr}(\mathbf R_i\,\mathbf \Xi)=0,
\quad i=1,\ldots,N_\mathrm S,\\
&\mathbf\Xi\succeq\mathbf0.
\end{aligned}
\label{eq_Fmin_sdr}
\end{equation}
The problem in \eqref{eq_Fmin_sdr} is a convex SDP and can be solved with standard convex optimization solvers. Because every feasible point of the exact rank-constrained formulation in \eqref{eq_Fmin_rank} is feasible for the relaxed SDP in \eqref{eq_Fmin_sdr}, the optimal value of \eqref{eq_Fmin_sdr} is an upper bound on the largest achievable target-mode strength among wavefronts with fidelity at least \(F_\mathrm{min}\). As demonstrated in Appendix~\ref{app_ambiguity_all}, this bound is agnostic to inevitable ambiguities in experimentally estimated proxy MNT parameters.

To trace out the bound on the strength--fidelity Pareto frontier, we sweep through a range of $F_\mathrm{min}$ values and solve the corresponding \eqref{eq_Fmin_sdr} for each of them. This approach is complementary to the one in Sec.~\ref{subsec_smin_sweep}; comparison of the two Pareto-frontier bounds can serve as a consistency check.

\section{Discrete Optimization Benchmarks}
\label{sec_discrete_optimizations}

To probe the tightness of the SDR-based bounds on the achievable wavefront-synthesis performance derived in Sec.~\ref{sec_Theory}, we compare them with the performance achieved by feasible designs obtained from several standard discrete-optimization techniques. We emphasize that we do \textit{not} propose these algorithms as new optimization methods; instead, we only use them as simple benchmarks to see how closely standard discrete-optimization techniques can approach the SDR-based bounds.

The considered discrete-optimization techniques differ only in how they generate candidate binary load vectors. For a given candidate \(\mathbf r\), we use the same evaluation procedure in all cases: we compute \(\mathbf H(\mathbf r)\) from \eqref{eq_MNT} and then globally solve the remaining input-wavefront optimization, as detailed in Appendix~\ref{app_fixed_load_input_optimization}. Consequently, any suboptimality of the reported feasible designs stems from the discrete search over binary load vectors, not from the continuous input-wavefront optimization for a fixed load vector.

\textit{Projected SDR solution (P-SDR):}

Although the optimizer of a relaxed SDP is generally not rank one and therefore does not directly yield a feasible binary load vector, it can be converted into candidate configurations. Our starting point is the optimized lifted matrix of the corresponding SDR, namely the optimizer of \(\tilde{\mathbf \Xi}\) for \eqref{eq_shape_sdr}, of \(\mathbf\Xi\) for \eqref{eq_strength_only_sdr} and \eqref{eq_Fmin_sdr}, and of \(\breve{\mathbf\Xi}\) for \eqref{eq_Smin_sdr}. We extract its leading eigenvector and, in addition, draw 80 complex Gaussian candidate vectors whose covariance is the optimized lifted matrix. This randomization exploits information contained in all significant eigendirections of the relaxed solution, rather than only its dominant eigenvector. For each of the resulting 81 candidates, we compute \(\mathbf z=\mathbf\Gamma\mathbf v+\mathbf B\mathbf x\) and assign the \(i\)th load to the binary state whose corresponding residual is smaller, i.e., we choose between \(\alpha\) and \(\beta\) by comparing \(|v_i-\alpha z_i|\) and \(|v_i-\beta z_i|\). We then evaluate the resulting binary configurations with the fixed-load input-wavefront optimization described in Appendix~\ref{app_fixed_load_input_optimization} and retain the best one.

\textit{Coordinate descent (CD):}
To initialize the CD, we evaluate the two uniform configurations, the projected-SDR candidate, and 80 randomly drawn binary configurations. Then, we start the CD sweeps from the best of these 83 candidates. We define one CD sweep as follows: for each tunable element in turn, we flip its binary state, re-evaluate the resulting configuration, and keep the flip whenever it improves the relevant objective. Since a single-bit flip only results in a rank-one change of the matrix to be inverted in \eqref{eq_MNT}, we efficiently evaluated tentative flips using the Woodbury identity~\cite{prod2023efficient} rather than by recomputing the full inverse from scratch. We perform at most 150 sweeps and stop earlier if a full sweep finds no improving single-bit flip or if the relative improvement over one sweep falls below \(10^{-10}\).

\textit{Genetic algorithm (GA):}
Our GA fitness function is the negative of the relevant optimization objective. For the optimization problems with minimum-strength or minimum-fidelity thresholds, we penalize infeasible configurations according to their threshold violation. Specifically, we add a large positive offset to the fitness of infeasible configurations, together with the corresponding threshold deficit. Our initial population contains the projected-SDR candidate, the two uniform configurations, and 57 random binary configurations. We use 40 generations and an elite count of 3, meaning that we copy the three best configurations unchanged to the next generation.

\textit{Exhaustive search (ES):}
For sufficiently small \(N_\mathrm S\), we exhaustively enumerate and evaluate all \(2^{N_\mathrm S}\) binary load configurations to identify the globally optimal one. To make this benchmark computationally tractable up to \(N_\mathrm S = 20\), we enumerate the configurations in Gray-code order, so that successive configurations differ by only one load state. We then evaluate the corresponding single-load update of the MNT inverse in \eqref{eq_MNT} using the Woodbury identity rather than recomputing the full inverse from scratch~\cite{prod2023efficient}.

\section{Experimental Results}
\label{sec_ExpResults}

\begin{figure*}
    \centering
    \includegraphics[width=2\columnwidth]{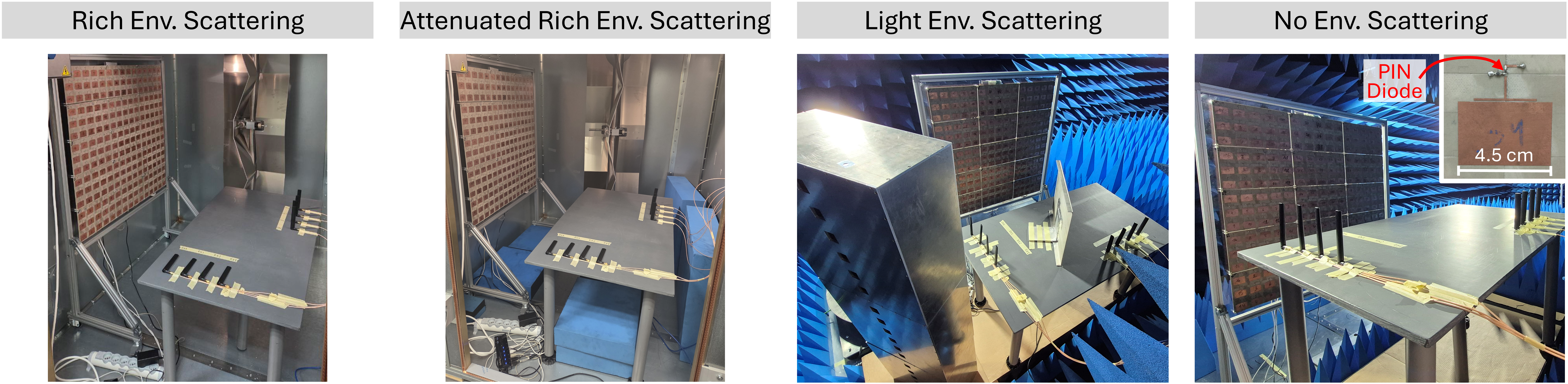}
    \caption{Photographic images of the four considered RIS-parametrized radio environments.}
    \label{Fig1}
\end{figure*}

\subsection{Experimental Setup and Proxy-Parameter Estimation}
\label{subsec_exp_setup}

We evaluate the proposed bounds using four experimental RIS-parametrized MIMO systems. All four systems use the same antenna and RIS hardware, but differ in the amount of environmental scattering. Although the same RIS is used in all four setups, the overall mutual coupling between the RIS elements varies substantially because environmental scattering strongly affects the inter-element coupling, as already experimentally shown in~\cite{rabault2023tacit,del2025experimentalreducedrank}. 
The transmitter and receiver each comprise four parallel, half-wavelength-spaced Wi-Fi antennas. The RIS prototype contains 225 half-wavelength-sized 1-bit-programmable elements designed for operation around \(2.45~\mathrm{GHz}\), of which we use up to 100 in this work; we keep the remaining RIS elements in a fixed reference state. The RIS elements are controlled through electrically small PIN diodes, in line with the lumped-load assumption underlying the multiport-network model~\cite{largeRIS_TCOM,del2025frozen}; details of the RIS design can be found in~\cite{KDF14,ahmed2025over}. The four considered radio environments are depicted in Fig.~\ref{Fig1}: (i) a reverberation chamber with rich environmental scattering, (ii) the same reverberation chamber loaded with absorbing material, (iii) an anechoic chamber containing a few metallic scattering objects, and (iv) the same anechoic chamber without additional scattering objects. The mode stirrer in the reverberation chamber is kept static throughout the measurements. 

We conduct measurements of the systems' \(4\times4\) end-to-end channel matrices with an eight-port vector network analyzer (VNA). Directly measuring the underlying MNT parameters is not possible in this setup. \textit{First}, the RIS has an integrated design: the PIN diodes are embedded in the RIS elements and the corresponding ``virtual'' ports are not connectorized. \textit{Second}, the full multiport system would have \(N=N_\mathrm T+N_\mathrm R+N_\mathrm S=108\) ports, far exceeding the number of available VNA ports. Moreover, a full-wave simulation of the complete experimental setup would be impractical because the detailed geometry and material composition of the radio environment are not fully known and the electrical size of the system is very large. We thus use experimentally estimated proxy MNT parameters. For each of the four environments, we estimate a proxy MNT parameter set from measured end-to-end channel matrices for known RIS control vectors using the approach of~\cite{ContRIS_LWC}. Such proxy parameters are not unique: different parameter sets can be operationally equivalent, in the sense that they yield the same mapping from RIS control vector to end-to-end channel matrix. The relevant ambiguity classes and the insensitivity of our bounds to these ambiguities are explained in Appendix~\ref{app_ambiguity_all}. Hence, any proxy MNT parameter set capable of accurate end-to-end channel predictions is suitable to evaluate our bounds. 
We quantify the predictive precision of each proxy MNT model by the metric \(\zeta\) used in~\cite{ContRIS_LWC}, which is defined analogously to a signal-to-noise ratio by treating the discrepancy between measured and model-predicted end-to-end channels for previously unseen RIS configurations as ``noise''. The obtained values are (i) 47.4~dB, (ii) 55.3~dB, (iii) 39.1~dB, and (iv) 42.7~dB. The resulting precisions correspond to relative prediction errors at approximately the percent level or below, even though the proxy model contains thousands of internal parameters, including \(N_\mathrm S(N_\mathrm S+1)/2=5050\) entries associated with inter-element coupling.

In the following analysis, we work with \(N_\mathrm{T}=2\) and \(N_\mathrm{R}=4\). This choice with $N_\mathrm{T}<N_\mathrm{R}$ avoids that the strength-agnostic wavefront-fidelity optimization becomes trivial, as discussed in Sec.~\ref{subsec_Metrics}. Since our experimentally calibrated proxy MNT model involves four transmitting antennas, we assume that the two unused transmitting antennas are terminated in matched loads. Since matched terminations have zero reflection coefficient, the corresponding reduction of the proxy MNT model simply amounts to selecting the corresponding subblocks of the experimentally calibrated proxy MNT model. Moreover, we consider different values of \(N_\mathrm{S}\le 100\) to examine the influence of the number of programmable RIS elements; in particular, we consider \(N_\mathrm{S}=20\) because exhaustive search is still feasible at that size. Whenever \(N_\mathrm{S}<100\), we fix the tunable loads associated with unused RIS elements to a reference state. In general, the corresponding reduction of the proxy MNT model would require a Schur-complement evaluation analogous to~\eqref{eq_MNT}. While the physical reflection coefficient of the load associated with the PIN diode is not zero in the reference state, our specific choice of proxy MNT parameterization represents the load's reference state by a zero load reflection (see details in~\cite{ContRIS_LWC}). Consequently, the Schur-complement evaluation simplifies to selecting the corresponding subblocks of the experimentally calibrated proxy MNT model.

\subsection{Experimental Results for Bounds on Wavefront-Synthesis Performance}
\label{subsec_ExpResults}

We solve all SDPs required to compute the reported SDR-based bounds using the SeDuMi solver within the CVX framework~\cite{cvx}.

\begin{figure*}[ht]
    \centering
    \includegraphics[width=2\columnwidth]{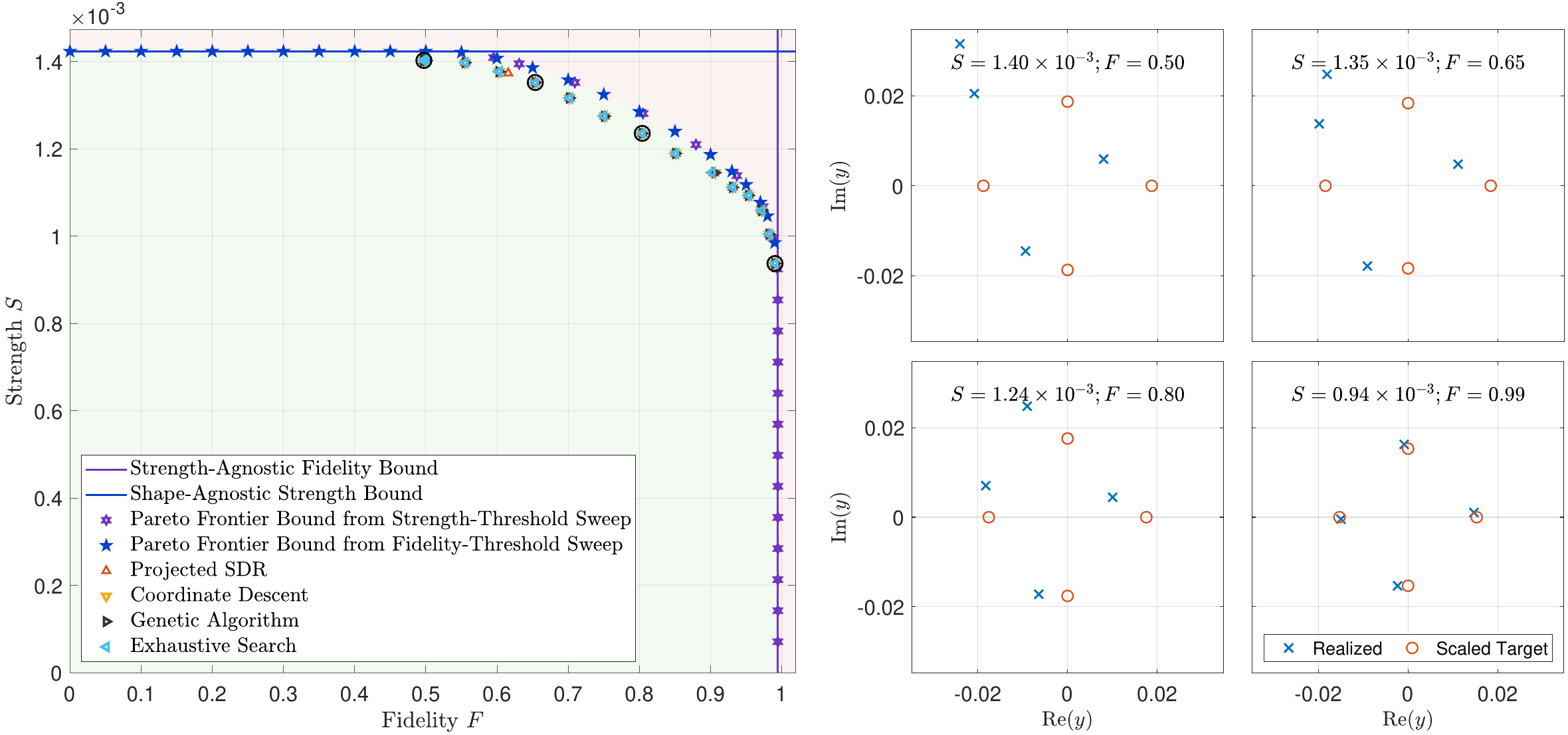}
    \caption{
    Strength--fidelity tradeoff and representative feasible wavefronts for targeting the phase-step wavefront \(\mathbf y_\star=[1,-\jmath,-1,\jmath]^\top/2\) in the rich-scattering environment with \(N_\mathrm{S}=20\). Left: SDR-based bounds and feasible discrete-optimization outcomes in the \((F,S)\) plane. The solid vertical and horizontal lines show the strength-agnostic fidelity bound and the shape-agnostic strength bound, respectively. Star markers show the bounds on the Pareto frontier obtained from the strength-threshold and fidelity-threshold sweeps. Open markers show feasible designs found by projected SDR, coordinate descent, genetic algorithm, and exhaustive search. The red shaded region is certified unattainable by at least one bound. Black circles mark the four feasible designs shown on the right. Right: realized output wavefronts (crosses) and optimally globally scaled target wavefronts (circles).
    }
    \label{Fig2}
\end{figure*}

For concreteness, we begin by considering the rich-scattering radio environment with \(N_\mathrm{S}=20\), targeting a ``phase-step'' wavefront: \(\mathbf y_\star^\mathrm{PhaseStep} = [1,\,-\jmath,\,-1,\,\jmath]^\top/2\). The corresponding results are displayed in Fig.~\ref{Fig2}. The strength-agnostic fidelity bound is \(99.5\,\%\), thus not excluding the possibility that the considered system can be configured to almost precisely generate the desired output-wavefront shape. The shape-agnostic strength bound of \(1.42\times 10^{-3}\) provides an upper bound on the achievable target-mode strength irrespective of the resulting wavefront shape. The two threshold-sweep constructions yield mutually consistent bounds on the strength--fidelity Pareto frontier: the points obtained from the strength-threshold and fidelity-threshold sweeps lie on the same smooth tradeoff curve. This agreement is reassuring because the two sweeps are complementary relaxations of the same underlying Pareto-frontier problem. Moreover, in the low-threshold limits, the Pareto-frontier bounds consistently reduce to the corresponding one-dimensional bounds: the fidelity-threshold sweep approaches the shape-agnostic strength bound for small \(F_\mathrm{min}\), whereas the strength-threshold sweep approaches the strength-agnostic fidelity bound for small \(S_\mathrm{min}\).

Since \(N_\mathrm{S}=20\), exhaustive search (ES) is feasible and provides the globally optimal discrete benchmark for the considered binary feasibility set. It therefore directly probes the tightness of our SDR-based bounds. We observe in Fig.~\ref{Fig2} that the ES results approach the bounds very closely. Without a strength constraint, the globally best achievable fidelity is \(99.1\,\%\), only \(0.4\) percentage points below our strength-agnostic fidelity bound. Similarly, without a fidelity constraint, the globally best achievable target-mode strength is \(1.40\times10^{-3}\), within \(1.4\,\%\) of our corresponding shape-agnostic strength bound. Across the full tradeoff, the ES-identified nondominated strength--fidelity points remain close to the Pareto-frontier bound, confirming that our bound is tight for this example. The other discrete optimization methods (P-SDR, CD, GA) achieve very similar performance to ES. This indicates that they are effective for this example, although ES remains the relevant reference for assessing bound tightness because it certifies global optimality over all binary configurations. 

For four selected nondominated ES-identified tradeoff points, marked by black circles in the left panel of Fig.~\ref{Fig2}, we display the corresponding realized and globally scaled target wavefronts in the right panels of Fig.~\ref{Fig2}. These panels illustrate the strength--fidelity tradeoff in the complex plane. Moving from the maximum-strength point to the maximum-fidelity point increases the fidelity from about \(50\,\%\) to about \(99\,\%\), but reduces the target-mode strength from \(1.40\times10^{-3}\) to \(0.94\times10^{-3}\), corresponding to a strength loss of approximately \(32.9\,\%\). Conversely, operating at maximum strength requires accepting a much lower fidelity, illustrating that both metrics are needed to characterize wavefront-synthesis performance.

Next, we examine the strength--fidelity plane more broadly across all four experimental setups, for \(N_\mathrm{S}\in\{20,100\}\), and for two target wavefronts: the ``phase-step'' target \(\mathbf y_\star^\mathrm{PhaseStep} \) already considered in Fig.~\ref{Fig2} as well as the following fixed random complex target: \(\mathbf y_\star^{\mathrm{rand}} = [0.36+0.24\jmath, 0.14+0.17\jmath, -0.75+0.36\jmath, -0.24+0.15\jmath]^{\top}\).
For \(N_\mathrm{S}>20\), exhaustive search is no longer computationally practical and is therefore omitted. Consequently, for these larger systems, we can make definitive statements about bound tightness only when the discrete optimization methods nearly attain the bound. When they remain separated from the bound, the observed gap cannot be unambiguously attributed: it may reflect suboptimality of the discrete optimization methods, looseness of the SDR-based bound, or both.

\begin{figure*}[ht]
    \centering
    \includegraphics[width=2\columnwidth]{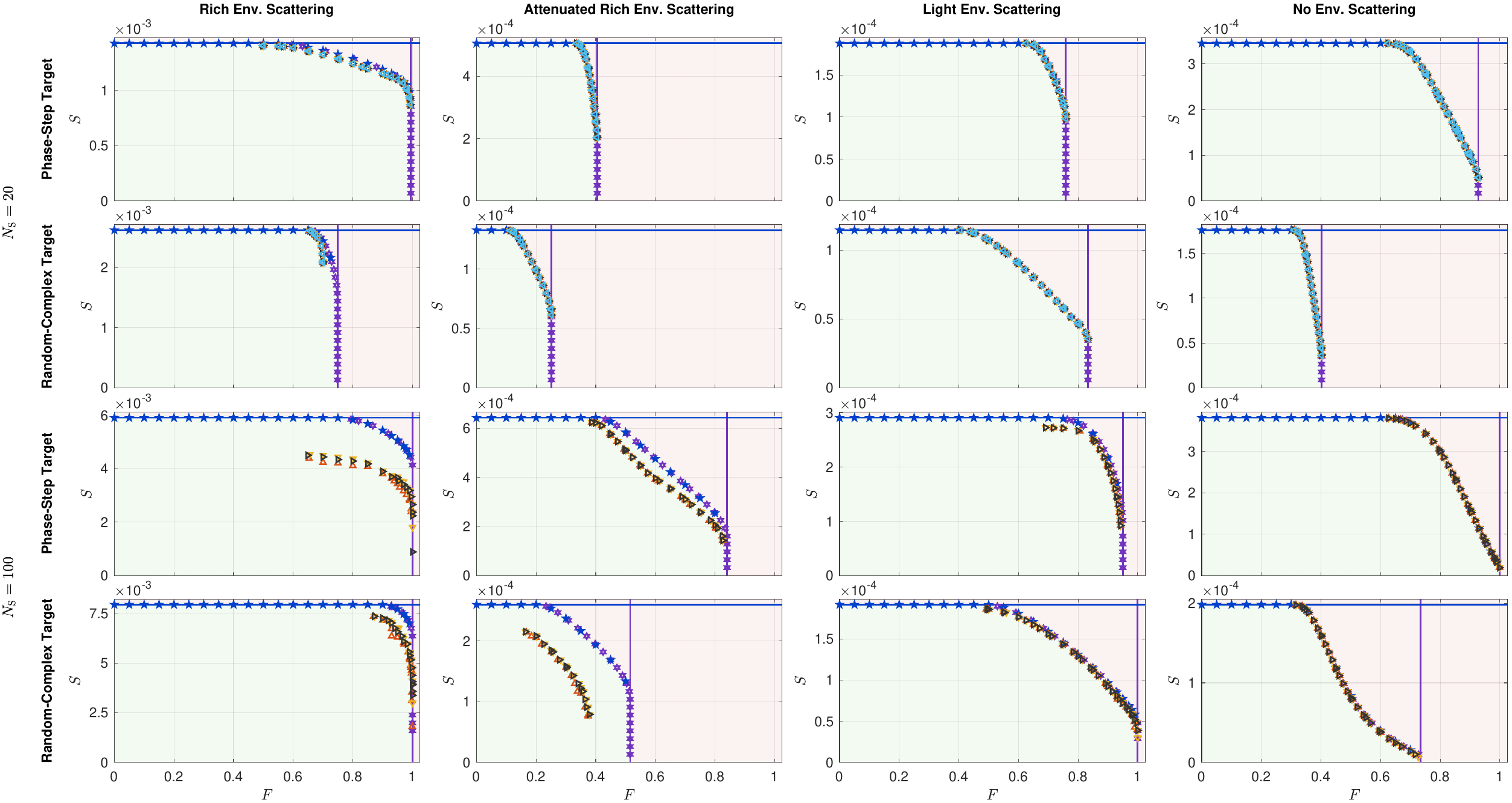}
    \caption{
Strength--fidelity tradeoff across the four experimental setups displayed in Fig.~\ref{Fig1} for \(N_\mathrm{S}\in\{20,100\}\) and for two distinct target wavefronts (\(\mathbf y_\star^\mathrm{PhaseStep} \) and $\mathbf y_\star^{\mathrm{rand}}$). The legend is the same as in the left panel of Fig.~\ref{Fig2}. Columns correspond to the four radio environments displayed in Fig.~\ref{Fig1}, while rows correspond to the considered combinations of \(N_\mathrm{S}\) and target wavefront. Exhaustive-search results are only available for $N_\mathrm{S} = 20$.}
    \label{Fig3}
\end{figure*}

The results in Fig.~\ref{Fig3} show that the SDR-based bounds and the feasible designs exhibit the same overall Pareto-frontier shape across a broad range of instances. Operationally, these bounds provide several definite practical insights: they certify unattainable fidelity values, unattainable target-mode strengths, and unattainable strength--fidelity combinations. For \(N_\mathrm{S}=20\), where exhaustive search is available, the feasible Pareto front is close to the SDR-based bound in all considered instances, confirming that the relaxation can be tight beyond the single example in Fig.~\ref{Fig2}. In the case of no environmental scattering, the identified feasible Pareto points lie on the SDR-based Pareto-frontier bounds even for \(N_\mathrm{S}=100\). This certifies the tightness of the bounds in these instances, despite the absence of exhaustive search. It also yields a further practical insight: whenever the feasible designs attain the bounds, no other optimization strategy can significantly outperform their performance.

Overall, the proximity between feasible designs and SDR-based bounds is less uniform in the other scattering environments. We observe a dependence on both the target wavefront and the location along the frontier. For example, in the rich-scattering environment with \(N_\mathrm{S}=20\), the phase-step target leads to feasible designs that very closely approach the fidelity side of the frontier, while a more visible, but still small, gap remains near the maximum-strength side. For the random-complex target, the behavior is reversed: the feasible designs more closely approach the maximum-strength side, while a larger gap remains near the high-fidelity side. For \(N_\mathrm{S}=100\), we observe larger gaps between the identified feasible points and the bounds in some cases. Since exhaustive search is unavailable for \(N_\mathrm{S}=100\), such gaps cannot be uniquely attributed to looseness of the SDR relaxation; they may also reflect the inability of the discrete optimizers to identify better binary configurations. We also note that the three heuristic discrete-optimization methods generally yield very similar performance for \(N_\mathrm{S}=100\).

\section{Discussion}
\label{sec_Discussion}

Our wavefront-synthesis bounds are broadly related to recent efforts on bounding the achievable performance of optical and nanophotonic devices~\cite{chao2022physical,miller2026fundamental}. Closest in spirit are bounds on multifunctional and tunable nanophotonic responses, where one structure is constrained to realize several prescribed responses~\cite{shim2024fundamental,gertler2025many}, and bounds on whether a static scatterer can implement a target linear input-output map at a specified fidelity~\cite{kuang2020computational}. These works are part of a growing research thrust on convex-relaxation-based electromagnetic and photonic-design bounds~\cite{molesky2020hierarchical,gustafsson2020upper,liska2021fundamental,zhang2021conservation,chao2022physical,amaolo2024maximum,virally2025many,miller2026fundamental}, to which the present work also belongs methodologically. An important difference is that a reconfigurable wave system realizes not a single wave response, but a family of physically admissible responses indexed by the control states. A second difference is that our bounds are prototype-aware, being evaluated for concrete prototypes of reconfigurable microwave systems based on experimentally calibrated physics-consistent proxy MNT models. Finally, to the best of our knowledge, bounds on the strength--fidelity Pareto frontier for target-wavefront synthesis have not previously been considered, neither for static nor reconfigurable systems, and neither in purely theoretical nor experimentally grounded form.

\section{Conclusion}
\label{sec_Conclusion}

To summarize, we established prototype-aware and electromagnetically consistent bounds on target-wavefront synthesis with PMs. We used a multiport-network description and SDRs of the resulting optimization problems to derive upper bounds on strength-agnostic fidelity, shape-agnostic target-mode strength, and the strength--fidelity Pareto frontier. We experimentally evaluated these bounds for four distinct RIS-parametrized MIMO setups, spanning rich-scattering to free-space-like radio environments, and showed that they certify unattainable performance regions while often being closely approached by feasible discrete-optimization benchmarks. Although we demonstrated the approach for a RIS prototype, it is not RIS-specific: because the multiport-network framework applies universally to lumped-element-reconfigurable wave systems, the same methodology extends directly to other PM embodiments, including DMAs. Our results establish experimentally grounded convex-relaxation bounds as practical benchmarks for assessing when a prototype is operated near its fundamental electromagnetic limit and when further optimization may still yield gains.

\appendices

\section{Insensitivity of the SDR Bounds to \\ Ambiguities in Proxy MNT Parameters}
\label{app_ambiguity_all}

In this Appendix, we demonstrate that the SDR-based bounds derived in \eqref{eq_shape_sdr}, \eqref{eq_strength_only_sdr}, \eqref{eq_Smin_sdr}, and \eqref{eq_Fmin_sdr} are invariant under the inevitable ambiguities of experimentally estimated proxy MNT parameters. 
First, we summarize three proxy-MNT ambiguity classes that relate operationally equivalent parameterizations of the same physical system. 
Second, we establish the common invariance mechanism used throughout this Appendix: each ambiguity induces an invertible linear change of variables that preserves the output wavefront, the input-power constraint, and the binary programmability constraints. 
Finally, we apply this mechanism separately to the four SDR-based bounds derived in Sec.~\ref{sec_Theory}. 

\subsection{Ambiguity Classes for Proxy MNT Parameter Sets}
\label{app_ambiguity_classes}

We denote proxy parameters with hats. A proxy MNT parameter set
\(\hat{\boldsymbol\theta}\triangleq
(\hat\alpha,\hat\beta,\hat{\mathbf H}_0,
\hat{\mathbf A},\hat{\mathbf\Gamma},\hat{\mathbf B})\)
is operationally equivalent to the true parameter set
\(\boldsymbol\theta\triangleq
(\alpha,\beta,\mathbf H_0,\mathbf A,\mathbf\Gamma,\mathbf B)\)
if it yields the same end-to-end mapping: \( \mathbf H(\mathbf u;\hat{\boldsymbol\theta})
= \mathbf H(\mathbf u;\boldsymbol\theta) \, \forall\,\mathbf u\in\mathbb B^{N_\mathrm S}, \) where \(\mathbf u\) is the control vector and the corresponding load vector is \( \mathbf r(\mathbf u)=\alpha\mathbf 1+(\beta-\alpha)\mathbf u \). The following three ambiguity classes generate valid proxy parameterizations~\cite{salmi2026electromagnetically,del2026electromagnetic}: 

\textit{Diagonal-similarity (DS) gauge:} 
\begin{equation}
\label{eq:gauge_DS_mimo}
\begin{aligned}
\hat{\mathbf H}_0 \;&=\; \mathbf H_0, 
&\qquad \hat\alpha \;&=\; \alpha,
&\qquad \hat\beta \;&=\; \beta,\\
\hat{\mathbf A} \;&=\; \mathbf A\mathbf D^{-1},
&\qquad \hat{\mathbf B} \;&=\; \mathbf D\mathbf B,
&\qquad \hat{\mathbf \Gamma} \;&=\; \mathbf D\mathbf \Gamma \mathbf D^{-1},
\end{aligned}
\end{equation}
where $\mathbf{D}\in \mathbb{C}^{N_\mathrm{S}\times N_\mathrm{S}}$ is an arbitrary invertible diagonal matrix.

\textit{Complex-scaling (CS) gauge:}
\begin{equation}
\label{eq:gauge_CS_mimo}
\begin{aligned}
\hat{\mathbf H}_0 \;&=\; \mathbf H_0, 
&\qquad \hat\alpha \;&=\; c\,\alpha,
&\qquad \hat\beta \;&=\; c\,\beta,\\
\hat{\mathbf A} \;&=\; \frac{1}{c}\mathbf A,
&\qquad \hat{\mathbf B} \;&=\; \mathbf B,
&\qquad \hat{\mathbf \Gamma} \;&=\; \frac{1}{c}\mathbf \Gamma,
\end{aligned}
\end{equation}
where \(c\in\mathbb C\setminus\{0\}\) is an arbitrary nonzero scalar.

\textit{M\"obius (M\"O) gauge:}
\begin{equation}
\label{eq:gauge_MO_mimo}
\begin{alignedat}{3}
\hat\alpha \;&=\; \mathcal M_m(\alpha),\quad
& \hat\beta \;&=\; \mathcal M_m(\beta),\quad
& \hat{\mathbf H}_0 \;&=\; \mathbf H_0 + m\,\mathbf A\mathbf F\mathbf B,\\
\hat{\mathbf A} \;&=\; k\,\mathbf A\mathbf F,\quad
& \hat{\mathbf B} \;&=\; k\,\mathbf F\mathbf B,\quad
& \hat{\mathbf \Gamma} \;&=\; (\mathbf \Gamma - m^*\mathbf I_{N_{\mathrm S}})\mathbf F,
\end{alignedat}
\end{equation}
where $\mathcal M_m(\rho)\triangleq (\rho-m)/(1-m^*\rho)$, $\mathbf F\triangleq (\mathbf I_{N_{\mathrm S}}-m\mathbf \Gamma)^{-1}$, $k\triangleq \sqrt{1-|m|^2}$, and \(m\in\mathbb C\) is an arbitrary scalar subject to the constraints that all inverses required to evaluate \eqref{eq:gauge_MO_mimo} exist, that $k\neq0$, and that $1-m^*\rho\neq 0$ for $\rho\in\{\alpha,\beta\}$.

\subsection{Common Invariance Mechanism}
\label{app_ambiguity_strategy}

The goal of this subsection is to show that each ambiguity transformation merely induces an invertible change of coordinates in the optimization variables. We first establish this change of coordinates at the vector level (i.e., in terms of $\bm\xi$), then show that it preserves the output wavefront, the input power, and the binary constraints, and finally lift the same change of coordinates to the SDP variables (i.e., in terms of $\bm\Xi$).

Throughout this subsection, we state the invariance mechanism for simplicity in terms of the generic stacked vector \(\bm\xi\) and its lifted matrix \(\mathbf\Xi=\bm\xi\,\bm\xi^\dagger\). The statements in this subsection can be directly applied to the normalized and Charnes--Cooper-scaled variables \(\tilde{\bm\xi}\), \(\breve{\bm\xi}\) and their lifted counterparts \(\tilde{\mathbf\Xi}\), \(\breve{\mathbf\Xi}\). Indeed, \(\tilde{\bm\xi}\) and \(\breve{\bm\xi}\) differ from \(\bm\xi\) only by scalar factors determined by \(\bm\xi^\dagger\mathbf Q_\mathrm D\bm\xi\), which is preserved under all considered ambiguity transformations. 

The bounds in Sec.~\ref{sec_Theory} are expressed in terms of the stacked variable \(\bm\xi=[\mathbf v^\top\ \mathbf x^\top]^\top\). In all three ambiguity classes, the transformed stacked variable satisfies \(\hat{\bm\xi}=\mathbf T\,\bm\xi\), where
\begin{equation}
\mathbf T
=
\begin{cases}
\begin{bmatrix}
\mathbf D & \mathbf 0\\
\mathbf 0 & \mathbf I_{N_\mathrm T}
\end{bmatrix},
& \text{DS},\\[4mm]
\begin{bmatrix}
c\mathbf I_{N_\mathrm S} & \mathbf 0\\
\mathbf 0 & \mathbf I_{N_\mathrm T}
\end{bmatrix},
& \text{CS},\\[4mm]
\begin{bmatrix}
k^{-1}(\mathbf I_{N_\mathrm S}-m\mathbf\Gamma) & -k^{-1}m\mathbf B\\
\mathbf 0 & \mathbf I_{N_\mathrm T}
\end{bmatrix},
& \text{M\"O},
\end{cases}
\label{eq_app_T_def}
\end{equation}
and \(\mathbf T\) is invertible in each admissible case.

The physical output wavefront is ambiguity-invariant by construction of the ambiguity classes. Indeed, with \(\mathbf C=[\mathbf A\ \mathbf H_0]\) and \(\hat{\mathbf C}=[\hat{\mathbf A}\ \hat{\mathbf H}_0]\), one verifies for all three ambiguity classes that
\begin{equation}
\hat{\mathbf C}\,\hat{\bm\xi}=\mathbf C\,\bm\xi
\quad
\Leftrightarrow
\quad
\hat{\mathbf C}\,\mathbf T=\mathbf C .
\label{eq_app_C_invariance}
\end{equation}
Since \(\mathbf T\) is invertible, we thus have \(\hat{\mathbf C}=\mathbf C\,\mathbf T^{-1}\). The definitions of \(\mathbf Q_\ell\), with \(\ell\in\{\mathrm S,\mathrm L,\mathrm D\}\), then yield
\begin{equation}
\hat{\mathbf Q}_\ell
=
\mathbf T^{-\dagger}\,\mathbf Q_\ell\,\mathbf T^{-1},
\qquad
\ell\in\{\mathrm S,\mathrm L,\mathrm D\}.
\label{eq_app_Q_transform}
\end{equation}

Since the lower block of \(\hat{\bm\xi}\) is always \(\mathbf x\), the input-power quadratic form is also ambiguity-invariant:
\begin{equation}
\hat{\bm\xi}^\dagger\,\mathbf E\,\hat{\bm\xi}
=
\bm\xi^\dagger\,\mathbf E\,\bm\xi
\quad
\Leftrightarrow
\quad
\mathbf T^\dagger\,\mathbf E\,\mathbf T=\mathbf E .
\label{eq_app_E_invariance}
\end{equation}

Next, we check the binary constraints. 
We define \(\ell_{i,\rho}=v_i-\rho z_i\) such that the binary constraint of the \(i\)th tunable element is \(\ell_{i,\alpha}^*\ell_{i,\beta}=0\). In the transformed parameterization, we analogously define \(\hat{\ell}_{i,\hat\rho}=\hat v_i-\hat\rho\,\hat z_i\).
\begin{itemize}
    \item For DS, \(\hat\rho=\rho\), \(\hat{\mathbf z}=\mathbf D\,\mathbf z\), and \(\hat{\ell}_{i,\hat\rho}=d_i\,\ell_{i,\rho}\).
    \item For CS, \(\hat\rho=c\,\rho\), \(\hat{\mathbf z}=\mathbf z\), and \(\hat{\ell}_{i,\hat{\rho}}=c\,\ell_{i,\rho}\).
    \item For M\"O, \(\hat\rho=\mathcal M_m(\rho)\), \(\hat{\mathbf v}=k^{-1}(\mathbf v-m\mathbf z)\), \(\hat{\mathbf z}=k^{-1}(\mathbf z-m^*\mathbf v)\), and hence \(\hat{\ell}_{i,\hat{\rho}} = \frac{k}{1-m^*\rho}\,\ell_{i,\rho}\).
\end{itemize}
Thus, in all three cases, each transformed binary factor equals the corresponding original binary factor multiplied by a nonzero scalar. Therefore, \(\hat{\ell}_{i,\hat\alpha}^*\hat{\ell}_{i,\hat\beta}=0\) if and only if \(\ell_{i,\alpha}^*\ell_{i,\beta}=0\).

We now translate these vector-level invariances to the lifted SDP variables. In the exact lifted formulation, \(\mathbf\Xi=\bm\xi\,\bm\xi^\dagger\). Hence, if \(\hat{\bm\xi}=\mathbf T\,\bm\xi\), the corresponding transformed lifted variable is
\begin{equation}
\hat{\mathbf\Xi}
=
\hat{\bm\xi}\,\hat{\bm\xi}^\dagger
=
\mathbf T\,\mathbf\Xi\,\mathbf T^\dagger .
\label{eq_congruenceTransform}
\end{equation}
We use the same congruence transformation for the relaxed SDR variable.

The map \(\mathbf\Xi\mapsto\hat{\mathbf\Xi}=\mathbf T\,\mathbf\Xi\,\mathbf T^\dagger\) preserves all lifted quadratic quantities associated with target-mode strength, leakage power, total output power, and input power. Indeed, combining it with \eqref{eq_app_Q_transform} gives
\begin{equation}
\mathrm{tr}(\hat{\mathbf Q}_\ell\,\hat{\mathbf\Xi})
=
\mathrm{tr}(\mathbf Q_\ell\,\mathbf\Xi),
\qquad
\ell\in\{\mathrm S,\mathrm L,\mathrm D\}.
\end{equation}
Similarly, combining \(\mathbf\Xi\mapsto\hat{\mathbf\Xi}=\mathbf T\,\mathbf\Xi\,\mathbf T^\dagger\) with \eqref{eq_app_E_invariance}, we obtain
\begin{equation}
\mathrm{tr}(\mathbf E\,\hat{\mathbf\Xi})
=
\mathrm{tr}(\mathbf E\,\mathbf\Xi).
\end{equation}

It remains to verify the preservation of the lifted binary constraints. From the factor-level relations above, for each ambiguity class there exist nonzero scalars \(\eta_{i,\rho}\) such that
\begin{equation}
\hat{\ell}_{i,\hat\rho}
=
\eta_{i,\rho}\,\ell_{i,\rho},
\qquad
\rho\in\{\alpha,\beta\}.
\end{equation}
Equivalently,
\begin{equation}
\hat{\mathbf g}_{i,\hat\rho}^{\top}\mathbf T
=
\eta_{i,\rho}\,\mathbf g_{i,\rho}^{\top},
\qquad
\rho\in\{\alpha,\beta\}.
\label{eq45}
\end{equation}
Multiplying the conjugate transpose of \eqref{eq45} with \eqref{eq45} gives
\begin{equation}
\mathbf T^\dagger\,
\hat{\mathbf g}_{i,\hat\alpha}^*
\hat{\mathbf g}_{i,\hat\beta}^{\top}\,
\mathbf T
=
\eta_{i,\alpha}^*\,\eta_{i,\beta}\,
\mathbf g_{i,\alpha}^*\,
\mathbf g_{i,\beta}^{\top}.
\label{eq55}
\end{equation}
Inserting the definitions of \(\mathbf R_i\) and \(\hat{\mathbf R}_i\) into \eqref{eq55} yields
\begin{equation}
\mathbf T^\dagger\,\hat{\mathbf R}_i\,\mathbf T
=
\eta_{i,\alpha}^*\,\eta_{i,\beta}\,\mathbf R_i .
\end{equation}
Therefore,
\begin{equation}
\mathrm{tr}(\hat{\mathbf R}_i\,\hat{\mathbf\Xi})
=
\mathrm{tr}(\mathbf T^\dagger\,\hat{\mathbf R}_i\,\mathbf T\,\mathbf\Xi)
=
\eta_{i,\alpha}^*\,\eta_{i,\beta}\,
\mathrm{tr}(\mathbf R_i\,\mathbf\Xi).
\end{equation}
Since \(\eta_{i,\alpha}^*\,\eta_{i,\beta}\neq0\), the constraint \(\mathrm{tr}(\mathbf R_i\,\mathbf\Xi)=0\) holds if and only if \(\mathrm{tr}(\hat{\mathbf R}_i\,\hat{\mathbf\Xi})=0\). Thus, the relaxed lifted binary constraints are preserved exactly by the congruence transformation.

Moreover, the map \(\mathbf\Xi\mapsto\hat{\mathbf\Xi}=\mathbf T\,\mathbf\Xi\,\mathbf T^\dagger\) preserves positive semidefiniteness because, for any vector \(\mathbf a\), one has \(\mathbf a^\dagger\,\hat{\mathbf\Xi}\,\mathbf a=(\mathbf T^\dagger\,\mathbf a)^\dagger\,\mathbf\Xi\,(\mathbf T^\dagger\,\mathbf a)\ge0\) whenever \(\mathbf\Xi\succeq\mathbf0\). Since \(\mathbf T\) is invertible, the map is bijective. It also preserves rank because left- and right-multiplication by invertible matrices do not change rank; specifically, \(\mathrm{rank}(\hat{\mathbf\Xi})=\mathrm{rank}(\mathbf T\,\mathbf\Xi\,\mathbf T^\dagger)=\mathrm{rank}(\mathbf\Xi)\). Thus, if the rank-one constraint is retained in the exact lifted formulation, it is preserved by the same congruence transformation. 

Altogether, each feasible lifted point in one parameterization corresponds bijectively to exactly one feasible lifted point in the other parameterization, with the same values of all objective and constraint expressions relevant to the SDRs. This is the common reason why all four SDR bounds are ambiguity-insensitive.

\subsection{Strength-Agnostic Fidelity Bound}
\label{app_ambiguity_shape}

The strength-agnostic fidelity SDR in \eqref{eq_shape_sdr} uses the normalized lifted variable \(\tilde{\mathbf\Xi}\). Since the denominator \(\bm\xi^\dagger\,\mathbf Q_\mathrm D\,\bm\xi\) is invariant under the ambiguity transformations, the normalized variable transforms by the same congruence map: \(\hat{\tilde{\mathbf\Xi}}=\mathbf T\,\tilde{\mathbf\Xi}\,\mathbf T^\dagger\). The common invariance mechanism then preserves positive semidefiniteness, the binary constraints, the normalization \(\mathrm{tr}(\mathbf Q_\mathrm D\,\tilde{\mathbf\Xi})=1\), and the objective \(\mathrm{tr}(\mathbf Q_\mathrm S\,\tilde{\mathbf\Xi})\). Because the map is bijective, the feasible objective values of the original and transformed SDPs coincide, so the optimal value of \eqref{eq_shape_sdr} is ambiguity-invariant.

\subsection{Shape-Agnostic Strength Bound}
\label{app_ambiguity_strength}

The shape-agnostic strength SDR in \eqref{eq_strength_only_sdr} uses the unnormalized lifted variable \(\mathbf\Xi\), which transforms as \(\hat{\mathbf\Xi}=\mathbf T\,\mathbf\Xi\,\mathbf T^\dagger\). The common invariance mechanism preserves positive semidefiniteness, the binary constraints, the input-power constraint \(\mathrm{tr}(\mathbf E\,\mathbf\Xi)\le1\), and the objective \(\mathrm{tr}(\mathbf Q_\mathrm S\,\mathbf\Xi)\). Because the map is bijective, the feasible objective values of the original and transformed SDPs coincide, so the optimal value of \eqref{eq_strength_only_sdr} is ambiguity-invariant.

\subsection{Minimum-Strength Fidelity Bound}
\label{app_ambiguity_smin}

The minimum-strength fidelity SDR in \eqref{eq_Smin_sdr} uses the Charnes--Cooper-scaled lifted variable \(\breve{\mathbf\Xi}\) and the scalar \(t\). Under the ambiguity transformations, these variables transform as \(\hat{\breve{\mathbf\Xi}}=\mathbf T\,\breve{\mathbf\Xi}\,\mathbf T^\dagger\) and \(\hat t=t\). The common invariance mechanism preserves positive semidefiniteness, the binary constraints, the denominator normalization \(\mathrm{tr}(\mathbf Q_\mathrm D\,\breve{\mathbf\Xi})=1\), the minimum-strength constraint \(\mathrm{tr}(\mathbf Q_\mathrm S\,\breve{\mathbf\Xi})\ge S_\mathrm{min}t\), the scaled input-power constraint \(\mathrm{tr}(\mathbf E\,\breve{\mathbf\Xi})\le t\), and the objective \(\mathrm{tr}(\mathbf Q_\mathrm S\,\breve{\mathbf\Xi})\). Because the map is bijective for every fixed \(S_\mathrm{min}\), the feasible objective values of the original and transformed SDPs coincide. Hence the optimal value of \eqref{eq_Smin_sdr} is ambiguity-invariant for every fixed \(S_\mathrm{min}\), and the entire \(S_\mathrm{min}\)-sweep is ambiguity-invariant.

\subsection{Minimum-Fidelity Strength Bound}
\label{app_ambiguity_fmin}

The minimum-fidelity strength SDR in \eqref{eq_Fmin_sdr} uses the unnormalized lifted variable \(\mathbf\Xi\), which transforms as \(\hat{\mathbf\Xi}=\mathbf T\,\mathbf\Xi\,\mathbf T^\dagger\). The common invariance mechanism preserves positive semidefiniteness, the binary constraints, the input-power constraint \(\mathrm{tr}(\mathbf E\,\mathbf\Xi)\le1\), the objective \(\mathrm{tr}(\mathbf Q_\mathrm S\,\mathbf\Xi)\), and the fidelity-threshold constraint
\[
(1-F_\mathrm{min})\,\mathrm{tr}(\mathbf Q_\mathrm S\,\mathbf\Xi)
-
F_\mathrm{min}\,\mathrm{tr}(\mathbf Q_\mathrm L\,\mathbf\Xi)
\ge0 .
\]
Because the map is bijective for every fixed \(F_\mathrm{min}\), the feasible objective values of the original and transformed SDPs coincide. Hence the optimal value of \eqref{eq_Fmin_sdr} is ambiguity-invariant for every fixed \(F_\mathrm{min}\), and the entire \(F_\mathrm{min}\)-sweep is ambiguity-invariant.

\section{Global Input-Wavefront Optimization \\for a Fixed Load Vector}
\label{app_fixed_load_input_optimization}

For a fixed binary load vector \(\mathbf r\), the end-to-end matrix
\(\mathbf H=\mathbf H(\mathbf r)\) is fixed and the remaining optimization is
only over the input wavefront \(\mathbf x\). Defining $\mathbf h\triangleq \mathbf H^\dagger\,\hat{\mathbf y}_\star$, $\mathbf G\triangleq \mathbf H^\dagger\,\mathbf P_\perp\,\mathbf H$, and $\mathbf W\triangleq \mathbf H^\dagger\,\mathbf H$, we have
\begin{subequations}
\begin{align}
S(\mathbf x)
&= |\hat{\mathbf y}_\star^\dagger\,\mathbf H\,\mathbf x|^2
= |\mathbf h^\dagger\mathbf x|^2, \\
L(\mathbf x)
&= \mathbf x^\dagger\,\mathbf G\,\mathbf x, \\
F(\mathbf x)
&= \frac{S(\mathbf x)}{S(\mathbf x)+L(\mathbf x)}
= \frac{\mathbf x^\dagger\,\mathbf h\,\mathbf h^\dagger\,\mathbf x}
{\mathbf x^\dagger\,\mathbf W\,\mathbf x}.
\end{align}
\end{subequations}
The four input-wavefront optimizations considered in this paper are continuous low-dimensional problems that can be solved globally once \(\mathbf r\) is fixed.
Throughout this Appendix, we interpret matrix inverses as Moore--Penrose pseudoinverses whenever the relevant matrix is singular.

\textit{Strength-Agnostic Fidelity.}
If \(\mathbf h=\mathbf0\), no input wavefront couples to the target mode and the numerator of the fidelity vanishes; otherwise, the following characterizations apply.
The strength-agnostic fidelity problem is the generalized Rayleigh-quotient problem
\begin{equation}
\max_{\mathbf x\neq\mathbf0}
\frac{\mathbf x^\dagger\,\mathbf h\,\mathbf h^\dagger\,\mathbf x}
{\mathbf x^\dagger\,\mathbf W\,\mathbf x}.
\end{equation}
Thus, an optimal direction is the dominant generalized eigenvector satisfying $\mathbf h\,\mathbf h^\dagger\,\mathbf x = \lambda\,\mathbf W\,\mathbf x$. 
When \(\mathbf W\) is nonsingular, a globally optimal normalized solution is \(\mathbf x_\mathrm{fid} = \mathbf W^{-1}\, \mathbf h / \|\mathbf W^{-1}\,\mathbf h\|_2\). 

\textit{Shape-Agnostic Strength.}
The shape-agnostic strength problem is
\begin{equation}
\max_{\|\mathbf x\|_2\le1} |\mathbf h^\dagger\mathbf x|^2 .
\label{eq60}
\end{equation}
If \(\mathbf h\neq\mathbf0\), the Cauchy--Schwarz inequality shows that the globally optimal solution is the matched-filter input \(\mathbf x_\mathrm{str} = \mathbf h / \|\mathbf h\|_2\), yielding a strength of \(\|\mathbf h\|_2^2\). If \(\mathbf h=\mathbf0\), the achievable target-mode strength is zero.

\textit{Minimum-Strength Fidelity.}
The minimum-strength fidelity problem,
\begin{equation}
\max_{\|\mathbf x\|_2\le1}
F(\mathbf x)
\quad
\mathrm{s.t.}\quad
|\mathbf h^\dagger\,\mathbf x|^2\ge S_\mathrm{min},
\label{eq61}
\end{equation}
is infeasible if \(S_\mathrm{min}>\|\mathbf h\|_2^2\). Otherwise, since
\(F(\mathbf x)\) is invariant under any nonzero scaling of \(\mathbf x\), any feasible \(\mathbf x\) with \(\|\mathbf x\|_2<1\) can be normalized to unit norm without changing \(F(\mathbf x)\) such that we can restrict the search to \(\|\mathbf x\|_2=1\).

We first compute the normalized strength-agnostic fidelity optimizer
\(\mathbf x_\mathrm{fid}\). If \(|\mathbf h^\dagger\,\mathbf x_\mathrm{fid}|^2\ge S_\mathrm{min}\), then \(\mathbf x_\mathrm{fid}\) is globally optimal for \eqref{eq61}. Otherwise, within the unit-norm search, the strength constraint is active at the relevant optimum. Hence, the relevant optimum lies on the boundary \(|\mathbf h^\dagger\,\mathbf x|^2=S_\mathrm{min}\). On this boundary, maximizing \(F(\mathbf x)\) is equivalent to minimizing \(L(\mathbf x)\). We thus face the problem
\begin{equation}
\begin{aligned}
\min_{\mathbf x}\quad
& \mathbf x^\dagger\,\mathbf G\,\mathbf x\\
\mathrm{s.t.}\quad
& |\mathbf h^\dagger\,\mathbf x|^2=S_\mathrm{min},\\
& \|\mathbf x\|_2^2=1 .
\end{aligned}
\label{eq_min_strength_fid_boundary}
\end{equation}
Applying the Karush--Kuhn--Tucker (KKT) first-order stationarity condition~\cite[Ch.~5, Sec.~5.5.3]{boyd2004convex} to \eqref{eq_min_strength_fid_boundary} gives
\begin{equation}
(\mathbf G+\chi\,\mathbf I)\,\mathbf x
=\nu\,\mathbf h\,\mathbf h^\dagger\,\mathbf x ,
\end{equation}
where \(\chi\in\mathbb R\) and \(\nu\in\mathbb R\) are scalar multipliers. Since \(\mathbf h^\dagger\,\mathbf x\neq0\) whenever \(S_\mathrm{min}>0\), any nonzero stationary point for which \(\mathbf G+\chi\,\mathbf I\) is invertible has the direction of \((\mathbf G+\chi\,\mathbf I)^{-1}\,\mathbf h\).

The KKT condition has restricted any stationary point on the boundary \(|\mathbf h^\dagger\,\mathbf x|^2=S_\mathrm{min}\) to a one-parameter family of directions. Now, it remains to select the scalar parameter that enforces the minimum-strength constraint. To this end, we use the normalized parametrization
\begin{equation}
\mathbf x(\mu)=
\frac{(\mathbf G+\mu\mathbf I)^{-1}\mathbf h}
{\|(\mathbf G+\mu\mathbf I)^{-1}\mathbf h\|_2},
\qquad \mu\ge0 .
\label{eq_x_mu_active_strength}
\end{equation}
For \(\mu=0\), this direction gives the member of the family that most strongly suppresses leakage.
As \(\mu\) increases, the influence of the leakage matrix \(\mathbf G\) is monotonically weakened, and the target-mode strength increases monotonically.\footnote{Both the monotonicity of \(S(\mathbf x(\mu))\) and the fact that \(\mathbf x(\mu)\) is leakage-minimizing at fixed \(S(\mathbf x(\mu))\) follow by diagonalizing \(\mathbf G\) and applying the Cauchy--Schwarz inequality to \((\mathbf G+\mu\mathbf I)^{-1/2}\mathbf h\) and
\((\mathbf G+\mu\mathbf I)^{1/2}\mathbf x\).} In the limit \(\mu\to\infty\), \(\mathbf x(\mu)\to \mathbf h/\|\mathbf h\|_2\), which is the matched-filter input and maximizes the target-mode strength. 
Therefore, if the unconstrained fidelity optimizer does not satisfy the prescribed strength threshold but the problem is feasible, the globally optimal input wavefront is
\begin{equation}
\mathbf x_\mathrm{F}=
\frac{(\mathbf G+\mu_\mathrm F\mathbf I)^{-1}\mathbf h}
{\|(\mathbf G+\mu_\mathrm F\mathbf I)^{-1}\mathbf h\|_2},
\end{equation}
where \(\mu_\mathrm F\ge0\) is the selected value of the search parameter \(\mu\) such that $|\mathbf h^\dagger\,\mathbf x_\mathrm{F}|^2=S_\mathrm{min}$. In our numerical implementation, we find $\mu_\mathrm F$ by a one-dimensional bracketed search. $\mathbf x_\mathrm{F}$ globally maximizes the fidelity among all inputs satisfying the prescribed target-mode strength.

\textit{Minimum-Fidelity Strength.}
For \(F_\mathrm{min}=0\), the minimum-fidelity strength problem reduces to the shape-agnostic strength problem in \eqref{eq60}.
For \(F_\mathrm{min}>0\), the problem is infeasible if \(\mathbf h=\mathbf0\) or if \(F_\mathrm{min}\) exceeds the strength-agnostic fidelity characterized above. Otherwise, we define \(\gamma\triangleq (1-F_\mathrm{min})/F_\mathrm{min}\), so that \(F(\mathbf x)\ge F_\mathrm{min}\) is equivalent to \(\mathbf x^\dagger\,\mathbf G\,\mathbf x \le \gamma\,|\mathbf h^\dagger\,\mathbf x|^2\).
Thus, for \(F_\mathrm{min}>0\), the minimum-fidelity strength problem is
\begin{equation}
\max_{\|\mathbf x\|_2\le1}
|\mathbf h^\dagger\,\mathbf x|^2
\quad
\mathrm{s.t.}\quad
\mathbf x^\dagger\mathbf G\,\mathbf x
\le
\gamma\, |\mathbf h^\dagger\,\mathbf x|^2 .
\label{eq__66}
\end{equation}
If \(\mathbf h\neq\mathbf0\) and the matched-filter input \(\mathbf h/\|\mathbf h\|_2\) satisfies the fidelity constraint, then the matched-filter input is globally optimal for \eqref{eq__66}. Otherwise, the fidelity constraint is active. By the same KKT-stationarity argument as above, now applied to the active constraint \(\mathbf x^\dagger\,\mathbf G\,\mathbf x=\gamma\,|\mathbf h^\dagger\,\mathbf x|^2\), any nonzero stationary point has the direction of \((\mathbf I+\tau\,\mathbf G)^{-1}\,\mathbf h\), where $\tau\in\mathbb{R}$. Hence, the optimum is
\begin{equation}
\mathbf x_\mathrm{S}(\mu_\mathrm{S})=
\frac{(\mathbf I+\mu_\mathrm{S}\,\mathbf G)^{-1}\mathbf h}
{\|(\mathbf I+\mu_\mathrm{S}\,\mathbf G)^{-1}\mathbf h\|_2},
\qquad \mu_\mathrm{S}\ge0 ,
\end{equation}
with \(\mu_\mathrm{S}\) chosen such that $F(\mathbf x_\mathrm{S}(\mu_\mathrm{S}))=F_\mathrm{min}$. In our numerical implementation, this scalar equation is again solved by a one-dimensional bracketed search. The resulting input wavefront globally maximizes the target-mode strength under the required fidelity.

\section*{Acknowledgment}
P.d.H. acknowledges I.~Ahmed, F. Boutet, and C. Guitton who, under P.d.H.'s supervision, previously built the RIS prototype for the work presented in~\cite{ahmed2025over}. Moreover, P.d.H. acknowledges J.~Sol, who provided technical support for setting up the experiments at IETR's QOSC test facility (which is part of the CNRS RF-Net network).

\bibliographystyle{IEEEtran}

\providecommand{\noopsort}[1]{}\providecommand{\singleletter}[1]{#1}%

\end{document}